\documentclass[aps,pra,onecolumn,groupedaddress,floatfix,longbibliography,superscriptaddress,notitlepage]{revtex4-2}
\usepackage{amsmath}
\usepackage{amssymb}
\usepackage{amsfonts}
\usepackage{graphicx}
\usepackage{bbold}
\usepackage{bm}
\usepackage{cancel}
\usepackage{times,float}
\usepackage{graphicx}
\usepackage[usenames,dvipsnames,svgnames]{xcolor}
\usepackage{hyperref}
\hypersetup{colorlinks=true, linkcolor=NavyBlue, citecolor=PineGreen,urlcolor=cyan}
\usepackage{multirow}
\usepackage{braket}
\usepackage{float}
\usepackage{diagbox}
\usepackage{epsfig,subfig}
\usepackage{tabularx}
\usepackage{changepage}
\usepackage{soul}

\begin{document}

\title{ Testing the scalar sector of the Standard-Model Extension with neutron gravity experiments}

\author{C. A. Escobar}
\email{carlos.escobar@ciencias.unam.mx}
\affiliation{Departamento  de  F\'isica,  Universidad  Aut\'onoma  Metropolitana-Iztapalapa, San Rafael Atlixco 186, 09340 Ciudad de M\'{e}xico, M\'{e}xico}

\author{A. Mart\'{i}n-Ruiz}
\email{alberto.martin@nucleares.unam.mx}
\affiliation{Instituto de Ciencias Nucleares, Universidad Nacional Aut\'{o}noma de M\'{e}xico, 04510 Ciudad de M\'{e}xico, M\'{e}xico}

\author{A. M. Escobar-Ruiz}
\email{admau@xanum.uam.mx}
\affiliation{Departamento  de  F\'isica,  Universidad  Aut\'onoma  Metropolitana-Iztapalapa, San Rafael Atlixco 186, 09340 Ciudad de M\'{e}xico, M\'{e}xico}

\author{Rom\'an Linares}
\email{lirr@xanum.uam.mx}
\affiliation{Departamento  de  F\'isica,  Universidad  Aut\'onoma  Metropolitana-Iztapalapa, San Rafael Atlixco 186, 09340 Ciudad de M\'{e}xico, M\'{e}xico}

\begin{abstract}
{In the present study we analyse, within the scalar sector of the Standard-Model Extension (SME) framework, the influence of a spontaneous Lorentz symmetry breaking on gravitational quantum states of ultracold neutrons. The model is framed according to the laboratory conditions of the recent high-sensitivity GRANIT and $q$Bounce experiments. The high-precision data achieved in such experiments allow us to set bounds on the symmetry breaking parameters of the model. The effective Hamiltonian governing the neutron's motion along the axis of free fall is derived explicitly. It describes a particle in a gravitational field with an effective gravitational constant controlled non-trivially by the Lorentz-violating parameters. In particular, using the exact wave functions and the energy spectrum, we evaluate both the heights associated with the quantum states and the transition frequencies between neighborhoring quantum states. By comparing our theoretical results with those reported in the GRANIT and the $q$Bounce experiments, upper bounds on the Lorentz-violating parameters are determined.  We also consider for the first time the gravity-induced interference pattern in a COW-type experiment to test Lorenz-invariance. In this case, an upper bound for the parameters is established as well.} 
\end{abstract}

\maketitle

\section{Introduction}
\label{Intro1}

General Relativity (GR) and the Standard Model (SM) are the most successful theories in describing nature, and they have been rigorously and extensively confirmed by experiments at the energy scales they operate. One of the main challenges of modern physics is the search for a quantum theory of gravity that merges them into a unified framework operating at the Planck scale. Quantum gravity effects may be associated with the breaking of Lorentz symmetry, and be observable at low-energy scales, described by an effective field theory called the Standard-Model Extension (SME) \citep{Colladay_Kostelecky_1997, Colladay_Kostelecky_1998, Bailey_Kostelecky_2006, Kostelecky_Tasson_2011}. The SME include the pure-gravity and minimally coupled SM Lagrangians, plus all possible Lorentz-violating terms constructed by the coupling of constant coefficients with gravitational and SM fields. These coefficients arises from the vacuum expectation values of basic tensor fields belonging to a more fundamental theory, such as string theory \cite{Kostelecky_1989, Kostelecky_1991}, and act as a background, thus producing Lorentz invariance violation. 

To date there are no experimental signs of a departure from Lorentz invariance. Therefore, it is generally assumed that the LV coefficients are highly suppressed, thus producing only small modifications in physically measurable quantities.  The experimental searches of violation of Lorentz invariance are mostly conducted in comparing the predictions of the SME with that of the SM and GR. This allow us to impose upper bounds to the possible values for the LV coefficients for the different sectors of the SME. These constraints are compiled in Ref. \cite{Kostelecky_DataTables}. Interestingly, the SME predicts new effects which are absent or forbidden in the standard physics, namely, birefringence in vacuum \cite{Kostelecky_Mewes_2002, Bailey_Kostelecky_2004, Martin_2016,  Martin_2017}, vacuum \v{C}erenkov radiation \cite{Hohensee_2009, Lehnert_2004,  Lehnert_2004_2} and forbidden decays \cite{Kostelecky_Pickering_2003, Colladay_2017}. 

In recent years, the SME has appeared as a potential candidate to describe departures from emergent Lorentz symmetry in condensed matter systems such as topological phases and some unconventional superconductors.  In these cases, LV coefficients are not necessarily small, and they are determined by the underlying band structure of the material. Examples of this are the photon and fermion sectors of the SME. The former exhibits an interesting parallel with the electrodynamics of topological phases \cite{Martin_TI_2015, Martin_TI_2016, Martin_TI_2016_2, Martin_WS_2019}, and the latter is well-suited to describe the electronic bands in crystalline solids \cite{Grushin_2012, Gomez_2022, Kostelecky_2022}. This nontrivial deep connection between Lorentz symmetry in high-energy physics and emergent Lorentz symmetry in condensed matter  physics could has important implications in both directions. On the one hand, topological phases could represent potential test bed for open challenges in the SME framework. On the other hand, the SME offers a theoretical framework for the description, realization and even prediction of novel phases of matter.

Most of the known fundamental particles in the Standard Model have spin,  being the Higgs boson the only exception. However,  even for a spinful particle, a subset of Lorentz-violating effects are spin-independent and hence can be handled as though the particle had zero spin.  On the experimental side, these effects can be tested by using unpolarized beams of spinful particles, for instance. This idea opens the possibility to test also the scalar sector of the SME, recently introduced in Ref. \cite{Kostelecky-Edwards_2018}, which is a general effective scalar field theory in any spacetime dimension that contain explicit perturbative spin-independent Lorentz-violating operators of arbitrary mass dimension.  The scalar sector of the SME has been recently explored within the frameworks of Casimir physics \cite{Medel_2020, Martin-Ruiz_2020, Escobar_2020, Escobar-Ruiz_2021},  superconductivity \cite{Furtado_2021},  Bose-Einstein condensates \cite{Tian_2021} and thermodynamical behaviour \cite{Aguirre_2021, Filho_2021}. In this paper we suggest that neutron physics in the presence of the gravitational field serves as a test bed for the scalar sector of the SME. Being neutron a spinful particle, such experiments must be performed with unpolarized beams of them.

The recent advances in neutron physics provide a unique opportunity to study the interaction of a particle in a quantum state with a gravitational field, and importantly, provides a broad arena for the search of physics beyond the Standard Model, in particular, the breakdown of Lorentz invariance.  In this paper we shall consider the effects of Lorentz invariance violation, as described by the scalar sector of the SME, on different experiments involving unpolarized beam of neutrons in the presence of the Earth's gravitational field. Some of these experiments are those performed at the Institute Laue-Langevin (ILL) with ultracold neutrons (UCNs), these are known as GRANIT and $q$Bounce experiments.  On the one hand, the GRANIT experiments have shown that an intense beam of UCNs moving in Earth's gravity field does not bounce smoothly but at certain well-defined quantized heights, such as it is predicted by quantum mechanics \cite{Nesvizhevsky_2002}.  On the other hand, the $q$Bounce experiments are able to measure the transition frequency between gravitational quantum states by means of a gravitational resonant spectroscopy method \cite{Cronenberg_2018}. These experiments confirm the predictions of quantum theory without additional forces. Therefore,  the good agreement between theory and experiments can be turned into constraints onto additional forces. This is precisely what motivates the first part of this paper.  In the second part we suggest that a neutron interferometric device, as the pioneering one proposed by Colella, Overhauser and Werner in 1975 \cite{COW_1975}, may serves also as a test bed for Lorentz invariance violation. This device specializes in the measurement of gravitationally induced phase shift when a beam of neutrons is sent to a detector by different paths which lies at different gravitational potentials. 

The first part of this paper is addressed to the analysis of LV contributions to the energy spectrum and transition frequencies of the quantum gravitational states of ultracold neutrons.  For the sake of comparison with the GRANIT and $q$Bounce experiments, we frame our analysis according to the laboratory conditions under which they were carried out.  Since these experiments use a narrow beam of UCNs prepared with an adjustable horizontal velocity of a few m/s, we decouple the motion along the free-falling axis from that in the perpendicular plane by using a semiclassical wave packet, thus leaving us with an effective nonrelativistic Hamiltonian describing the quantum dynamics of free-falling UCNs.  We solve the Schr\"{o}dinger equation and determine in an analytical fashion the wave function and energy spectrum. Comparing our theoretical results with the experimental data of GRANIT and $q$Bounce collaborations, we set upper bounds for the Lorentz-violating coefficients of the scalar sector of the SME. Using the same Hamiltonian model,  and within the semiclassical prescription of matter-wave interferometry, in the second part of this paper we compute the gravitationally induced phase shift in a COW-type experiment and compare with the maximal experimental precision reached up to date.

The outline of this work is the following. In Section \ref{Model1a} we present the scalar LV model we deal with and some of its main properties.  The nonrelativistic Hamiltonian that describes the quantum dynamics of UCNs in the presence of the gravitational field is derived in Section \ref{Heffec1}.  In Section \ref{QGS-UCNs} we calculate the exact wave function and energy spectrum of the system.  Upon comparison of our exact theoretical results with data of the GRANIT and $q$Bounce experiments,  in Section \ref{ExpData} we set upper bounds on the LV parameters. Section \ref{COW-sect} considers neutron-interferometry as a possible test of Lorentz invariance violation, and comparing the theoretical results with the current precision in COW-type experiments, we get upper bounds for the LV parameters. Finally, in Section \ref{Conclu1} we conclude with a discussion of our results and future outlines.

\section{The model} \label{Model1a}

Consider a real scalar field $\Psi$ of mass $m$ in flat spacetime,  with the Minkowski metric $\eta _{\mu \nu}$ with signature $(+,-,-,-)$.  The effective Lagrange density describing the dynamics of $\Psi$ in the presence of Lorentz-violating effects can be written as \cite{Kostelecky-Edwards_2018}
\begin{align}
\mathcal{L} = \frac{1}{2} h ^{\mu \nu} \partial _{\mu} \Psi\,\partial _{\nu} \Psi - \frac{1}{2} \left( mc / \hbar \right) ^{2}  \Psi ^{2} , \label{LagrangianLV}
\end{align}
where $h ^{\mu \nu}$ is a symmetric second-rank tensor that represents a constant background, independent of the spacetime coordinates. Accordingly it does not transform as a tensor under {\em active} Lorentz transformations.  Naturally, some criteria as causality, a positive energy and stability impose restrictions on the components of $h ^{\mu \nu}$ \cite{Kosteleck_Lehnert_2001}.  

The Euler-Lagrange equations of motion for the theory (\ref{LagrangianLV}) are
\begin{align}
\left[ h _{\mu \nu} \partial ^{\mu} \partial ^{\nu} + \left( mc / \hbar \right) ^{2} \right] \Psi = 0 , \label{KleinGordonLV}
\end{align}
whilst the stress-energy tensor of the model takes the form
\begin{align}
T ^{\mu \nu} = h ^{\mu \alpha} \partial _{\alpha} \Psi \partial ^{\nu} \Psi - \eta ^{\mu \nu} \mathcal{L} .  \label{EnergyMomentumTensor}
\end{align}
Note that for a general $h ^{\mu \nu}$, unlike most of the standard cases where Lorentz symmetry is preserved, the tensor $T ^{\mu \nu}$ cannot be symmetrized because its antisymmetric part is no longer a total derivative.  As can be directly verified, the stress-energy tensor (\ref{EnergyMomentumTensor}) is conserved ($\partial _{\mu} T ^{\mu \nu} = 0$), but it is not traceless ($T _{\phantom{\mu} \mu} ^{\mu}\neq 0$).

The emergence of the $h^{\mu\nu}$-tensor can be understood from different approaches.  For example,  it arises from the vacuum expectation value of some tensor fields belonging to a more fundamental theory (such as loop quantum gravity and some string theories), thus specifying privileged space-time directions and implying the breakdown of Lorentz invariance.  To date there are no experimental signatures of a departure from Lorentz symmetry in quantum field theory and general relativity. Hence, in these frameworks, it is customary to take $\vert h ^{\mu \nu} \vert \ll 1$ in all earth-based frames. High-precision experimental measurements of the properties of particles have allowed to set bounds to the large set of parameters characterizing Lorentz-violating effects, all of them summarized in Ref. \cite{Kostelecky_DataTables}.

There are also different scenarios in which the theory (\ref{LagrangianLV}) may arise, and for which the different components of the $h^{\mu\nu}$-tensor are not necessarily small. For example, $h^{\mu\nu}$ can also represent a constant background field due to gravitational effects \cite{Chang_2012}, arising from the zeroth order approximation in a curved spacetime \cite{Zhang_2017}. In a more realistic framework, the theory (1) may arise in condensed matter systems, in which the fundamental symmetries are naturally broken. For example, the tensor $h^{\mu\nu}$ may contains the information regarding the optical properties of a ponderable media. In any case, this theory has attracted great attention recently and has motivated a number of investigations.

\section{Effective Hamiltonian}\label{Heffec1}

Quantum gravitational states of ultracold neutrons have been proposed as a potential handle to distinguish between Lorentz-invariant and Lorentz-violating theories.  For example,  Refs.  \cite{Martin_2018, Xiao_2020, Ivanov_2019} have used UCNs, quantized in the Earth's gravitational field, as a tool to probe effects of violation of Lorentz invariance within the fermion sector of the SME by using the experimental data on the energy levels of the lowest quantum gravitational states. In a similar fashion, UCNs have also been used as a tool for probing of beyond-Riemann gravity, in particular within the gravitational sector of the SME \cite{Escobar_2019, Ivanov_2021}. In every case, a nonrelativistic limit of the underlying relativistic theories was required, since the velocities of UCNs are quite small as compared with the velocity of light in vacuum. This motivates for the search, first, of the nonrelativistic limit of the equation of motion (\ref{KleinGordonLV}). To this end we use the usual ansatz factoring the oscillatory rest mass energy term: $\Psi ({\bf{r}},t) = \psi({\bf{r}},t) \, e ^{-\frac{i}{\hbar}mc ^{2} t } $, with $\psi({\bf{r}},t) = \phi _{E} ({\bf{r}})\, e ^{-\frac{i}{\hbar} E ^{\prime} t}$, being $E^{\prime} = E- mc^{2} \approx p^{2}/2m$ the kinetic energy.  In the nonrelativistic limit $E ^{\prime} \ll mc^{2}$ and $v=p/m \ll c$,  and hence we arrive to the following equation
\begin{align}
 i \hbar h_{00} \frac{\partial}{\partial t} \psi({\bf{r}},t) &= \left[ \frac{ \hbar ^{2} }{2m} h_{ij} \partial ^{i} \partial ^{j}   - i  \hbar c \, h _{0i} \partial ^{i} + \frac{1}{2} (1-h _{00}) m  c ^{2} \right] \psi({\bf{r}},t) .  \label{NonRelLimit}
\end{align}
The last constant term only shifts the energy spectrum, and hence no spectroscopic signature is left by its presence (i.e.  transition frequencies are unaffected).  So we can safely drop it from the above equation.  The resulting nonrelativistic Hamiltonian can be obtained from the equation of motion (\ref{NonRelLimit}) which, read in the eigenvalue form $i \hbar \partial _{t} \psi = \hat{\mathcal{H}} _{\textrm{free}} \psi$, yields
\begin{align}
\hat{\mathcal{H}} _{\textrm{free}} \equiv -\frac{1}{2mh_{00}} h_{ij} \hat{p} ^{i} \hat{p} ^{j} + \frac{1}{h _{00}} h _{0i}\hat{p} ^{i} c . \label{HamiltonianFree}
\end{align}
In the Lorentz-symmetric case, where $h_{\mu \nu} = \eta _{\mu \nu}$, the Hamiltonian (\ref{HamiltonianFree}) properly reduces to $\hat{H} = \frac{\hat{p} ^{2}}{2m}$, as expected. Therefore, the evolution of UCNs in such a theory is described by the effective Hamiltonian
\begin{align}
\hat{\mathcal{H}} _{\textrm{eff}} = \hat{\mathcal{H}} _{\textrm{free}} + mgz, \label{EffectiveHam}
\end{align}
where $mgz$ is the gravitational potential of the Earth with the standard gravitational acceleration $g$. The $z$-direction is chosen as the free falling axis.

The next goal is to adapt our analysis to the conditions under which experiments are carried out.  To this end we briefly recall both the GRANIT and $q$Bounce experiments performed at the Institute Laue-Langevin (ILL), Grenoble.  We shall discuss later the COW experiment. The experiments consist in allowing a narrow beam of UCNs generated by a source at the ILL reactor to fall towards a horizontal mirror due to the influence of the Earth's gravitational field.  An absorber/scatterer is placed above the first mirror at a certain height in order to select different quantum states. The neutrons are guided through this mirror-absorber-system in such a way that they are in first few quantum states. After this preparation process, the GRANIT experiments have been able to record that neutrons 
no longer move continuously in the vertical direction, but rather jump from one height to another, as predicted by quantum mechanics.  On the other hand,  the $q$Bounce collaboration developed a gravitational resonant spectroscopy method, which allows to measure the transition frequency between gravitational quantum states. Such transitions between quantum states in gravity above the mirror are driven by vibrating the mirror surface. Guided by these experiments, for the purposes of this paper, we have to compute the LV contributions to the energy spectrum and transition frequencies of the quantum gravitational states of UCNs, induced by the effective Hamiltonian (\ref{EffectiveHam}).

Our choice of the laboratory frame is related to the above discussed experiments, performed at ILL in Grenoble. As mentioned in Ref.  \cite{Ivanov_2019}, we may neglect the Earth's rotation assuming that the ILL laboratory frame is an inertial one.  We assume the gravitational force to act in $z$-direction, while mirrors are aligned with the $xy$-plane.  The motion in $x$- and $y$-direction is free and completely decouples from that in $z$-direction. The beam of UCNs is prepared with an adjustable horizontal velocity of a few m/s. So, the horizontal motion obeys classical laws while the vertical motion is described by quantum theory. We may therefore model the horizontal motion by a Gaussian wave packet of the form \cite{Martin_2018} 
\begin{align}
\Psi (\mathbf{r} _{\perp} ) = \frac{1}{\sqrt{\pi} \sigma } e ^{ {\frac{i}{\hbar}} \mathbf{p} _{\perp} \cdot \mathbf{r} _{\perp} - \frac{\mathbf{r} _{\perp} ^{2} }{ 2 \sigma ^{2} } } ,  \label{AnsatzGaussian}
\end{align}
where $\mathbf{r} _{\perp} = (x,y)$ and $\mathbf{p} _{\perp} = (p_{x},p_{y})$ represent the transversal coordinates and momenta on the perpendicular plane to the free fall motion of the UCNs.  Here,  $\vert \mathbf{p} _{\perp} \vert \sim 24$ eV, such that the energy is $\sim 10^{-7}$ eV.  The characteristic width $\sigma$ of the wave packet must be taken very small due to the classical horizontal neutron's motion. The wave function in Eq. (\ref{AnsatzGaussian}) allows to derive a reduced one-dimensional Hamiltonian $\hat{\mathcal{H}}_{z}$ that governs the vertical $z$-motion as follows
\begin{align}
\hat{\mathcal{H}} _{z}  \equiv  \langle \hat{\mathcal{H}} _{\textrm{eff}} \rangle = \int \Psi ^{\ast} (\mathbf{r} _{\perp} ) \, \hat{\mathcal{H}} _{\textrm{eff}} \,\Psi (\mathbf{r} _{\perp)} \, d ^{2} \mathbf{r} _{\perp} . \label{Hz1} 
\end{align}
Since deviations from the Lorentz symmetric case are expected to be small, we write $h^{\mu\nu}=\eta^{\mu\nu}+\delta h^{\mu\nu}$, where $\delta h_{\mu\nu}$ is a small perturbation. That way, the effective Hamiltonian $\hat{\mathcal{H}}_{\textrm{eff}}$ (\ref{EffectiveHam}) takes the form
\begin{align}
\hat{\mathcal{H}} _{\textrm{eff}} & \ = \  \frac{1}{2m} (1 - \delta h _{00} ) \hat{p} ^{2} \ + \ m\,g\,z \ - \ \frac{1}{2m} \delta h _{ij} \hat{p} ^{i} \hat{p} ^{j}\ - \ \delta h _{0i} \,\hat{p} _{i}\, c \ ,
\end{align}
where we have neglected second order contributions on $\delta h _{\mu \nu}$.  To evaluate the corresponding average values we first split the  terms in $\hat{\mathcal{H}} _{\textrm{eff}}$ as follows. Quantities with latin indices of the middle of the alphabet ($i,j,k,l$) denote spatial components $x,y,z$; while quantities with latin indices from the beginning of the alphabet ($a,b,c,d$) refer only to the perpendicular neutron's motion coordinates $x,y$. Thus, adopting the above convention we have
\begin{align}
 \langle \hat{p} ^{2} \rangle  =  \langle \hat{p} _{a} \hat{p} _{a} \rangle  + \hat{p} _{z} ^{2} \, ,  \qquad  \langle \delta h _{0i} \hat{p} _{i} \rangle = \delta h _{0a} \langle \hat{p} _{a} \rangle + \delta h _{0z} \hat{p} _{z} ,
\end{align}
and
\begin{align}
\langle \delta h_{ij} \hat{p}_{i} \hat{p} _{j} \rangle = \delta h _{ab}   \langle \hat{p} _{a} \hat{p} _{b} \rangle + 2 \delta h _{za} \langle \hat{p} _{a} \rangle \hat{p} _{z} + \delta h _{zz} \hat{p} ^{2} _{z}  .
\end{align}
The expectation values $\langle \hat{p} _{a} \rangle$ and  $\langle \hat{p} _{a} \hat{p} _{b} \rangle$ over the wave function (\ref{AnsatzGaussian}) can be immediately obtained. Explicitly, they read
\begin{align}
\langle \hat{p} _{a} \rangle = p _{a} \, ,   \qquad  \langle\hat{p} _{a} \hat{p} _{b} \rangle =  p _{a} p _{b}  + \frac{\hbar ^{2}}{2 \sigma ^{2}} \delta _{ab} .  \label{SomeExpectations}
\end{align}
Eventually, the reduced one-dimensional Hamiltonian $\hat{\mathcal{H}} _{z}$ of Eq. (\ref{Hz1}) takes the final form
\begin{align}
\hat{\mathcal{H}} _{z} = E _{\perp} + \frac{1}{2m} \left(1 -  \delta h _{00} - \delta h _{zz} \right) \hat{p} ^{2} _{z} - \frac{1}{m} \left( \delta h _{za} p _{a} + \delta h _{0z} m c \right) \hat{p} _{z} + mgz  ,  \label{EffectiveHamiltonian-LV}
\end{align}
here 
\begin{align}
E _{\perp}  &= \frac{1}{2m}(1-\delta h _{00}) \left( p _{a} p _{a} + \frac{\hbar ^{2}}{ \sigma ^{2}} \right) - \frac{1}{2m}\delta h _{ab} \left( p _{a} p _{b} +  \frac{ \hbar ^{2}}{2 \sigma ^{2}} \delta _{ab}  \right) -  \delta h _{0a} p _{a}  c \ ,
\end{align} 
contains constant energy terms alone, which can be safely disregarded since they do not produce modifications on the energy eigenvalues nor on the transition frequencies. Hereafter, it will be omitted. The eigenvalue problem of the Hamiltonian (\ref{EffectiveHamiltonian-LV}) is treated in the next Section.

\section{Quantum gravitational states of UCNs in a LV background} \label{QGS-UCNs}

The main goal of this Section is to derive the exact quantum gravitational states and the energy spectrum of ultracold neutrons in the presence of a Lorentz-violating background. Afterwards, we compare these theoretical results with those obtained in the GRANIT and $q$Bounce experiments. By taking the maximal experimental uncertainties we will be able to establish upper bounds on some components of $\delta {h} _{\mu \nu}$. 

Let us consider the one-dimensional spectral problem for the Hamiltonian operator (\ref{EffectiveHamiltonian-LV}):
\begin{align}
\hat{\mathcal{H}} _{z}\, \psi (z)= \left[ \frac{\hat{p} _{z} ^{2}}{2 m} (1 - \xi ) - \frac{1}{m} \wp \, \hat{p} _{z} + mgz \right]\, \psi (z)= E \, \psi (z)  ,  \label{HamLV}
\end{align}
here $\xi \equiv \delta h _{00} + \delta h _{zz}$ is a dimensionless small LV parameter ($|\xi|\ll 1$) and $\wp \equiv \delta h _{za} p _{a} + \delta h _{0z} m c$ is also an LV coefficient with dimensions of momentum. As a first step, we make a gauge rotation of (\ref{EffectiveHamiltonian-LV}) to eliminate the linear in $\hat{p}_{z}$. Taking the gauge factor $\Gamma = e ^{i\mathfrak{p} z / \hbar }$ with
\begin{align}
\mathfrak{p} = \frac{\wp }{ 1- \xi }\ ,  \label{pFrak}
\end{align}
we construct the one-dimensional Schr\"odinger operator
\begin{equation}
\label{Hhz}
{\hat {h}} _{z}\ \equiv \ \Gamma^{-1}\,{ \hat {\mathcal H}}_{z}\,\Gamma =  \frac{\hat{p} _{z} ^{2}}{2 m} (1 - \xi ) + mgz = - (1-\xi ) \,\frac{\hbar ^{2}}{2\,m} \, \frac{d^2}{dz^2} + mgz   ,
\end{equation}
where without loss of generality the LV quadratic constant term $E _{0} = - \frac{\mathfrak{p} ^{2}}{2m} (1-\xi )$ was omitted. The spectra of ${\hat {H}} _{z}$ and ${\hat {h}} _{z}$ coincide whereas their eigenfunctions are related by the gauge factor $\Gamma$.
Next, in the corresponding spectral problem ${\hat {h}} _{z} \, \phi(z)=E \, \phi(z)$ we make the change of variable $z \rightarrow u\,l _{0}\,(1-\xi) ^{1/3}$ to obtain the equivalent dimensionless equation
\begin{equation}
\label{hz}
\left( - \frac{d ^{2}}{d u ^{2}} + u \right) \phi(u) = \epsilon\,\phi(u) , 
\end{equation}
where $\epsilon = \frac{E}{(1-\xi) ^{1/3}mgl_0}$ and $l _{0}=\sqrt[3]{\hbar ^{2}/(2m ^{2} g)}$ is the gravitational length. 

Now one can proceed in the usual manner to get $\phi(u)$ and $\epsilon$. The solution must obey the following boundary conditions: $\phi (u)$ must vanish asymptotically as $u \to \infty$, and $\phi (u=0)=0$ because of the presence of a mirror at $u = 0$. The general solution of the eigenvalue equation (\ref{hz}) can be written in terms of the Airy functions $\mbox{Ai}$ and $\mbox{Bi}$ \cite{Vallee_2004}. Since the latter goes to infinity as its argument grows, it is not a physically acceptable solution for the problem at hand. Therefore, we find the solution
\begin{equation}
\phi(u) = \mbox{Ai} ( u - \epsilon ) .
\end{equation} 
The energy spectrum is obtained from the boundary condition at the ground level $\phi (u=0)=0$. Thus,
\begin{align}
\epsilon_{n} =  - a _{n} \qquad ; \qquad n=1,2,3,\ldots \ , \label{EnergySpectrum}
\end{align}
being $a _{n}$ the $n$-th zero of the Airy function $\textrm{Ai}$. Accordingly, the exact normalized solution of (\ref{EffectiveHamiltonian-LV}) 
is given by
\begin{align}
\psi _{n}(z) = \frac{1}{\sqrt{(1-\xi) ^{1/3}l _{0}} \, \mbox{Ai} ^{\prime} ( a_{n} ) } e ^{i \mathfrak{p} z / \hbar } \mbox{Ai} \left[ a_{n} + \frac{z}{(1-\xi) ^{1/3}l _{0}} \right] , \label{Wave-Function-Fin}
\end{align}
with energy 
\begin{align}
E_{n} = - m g l _{0} (1-\xi ) ^{1/3} a _{n} \ . \label{EnergySpectrum}
\end{align}

\begin{figure}
\includegraphics[scale=0.5]{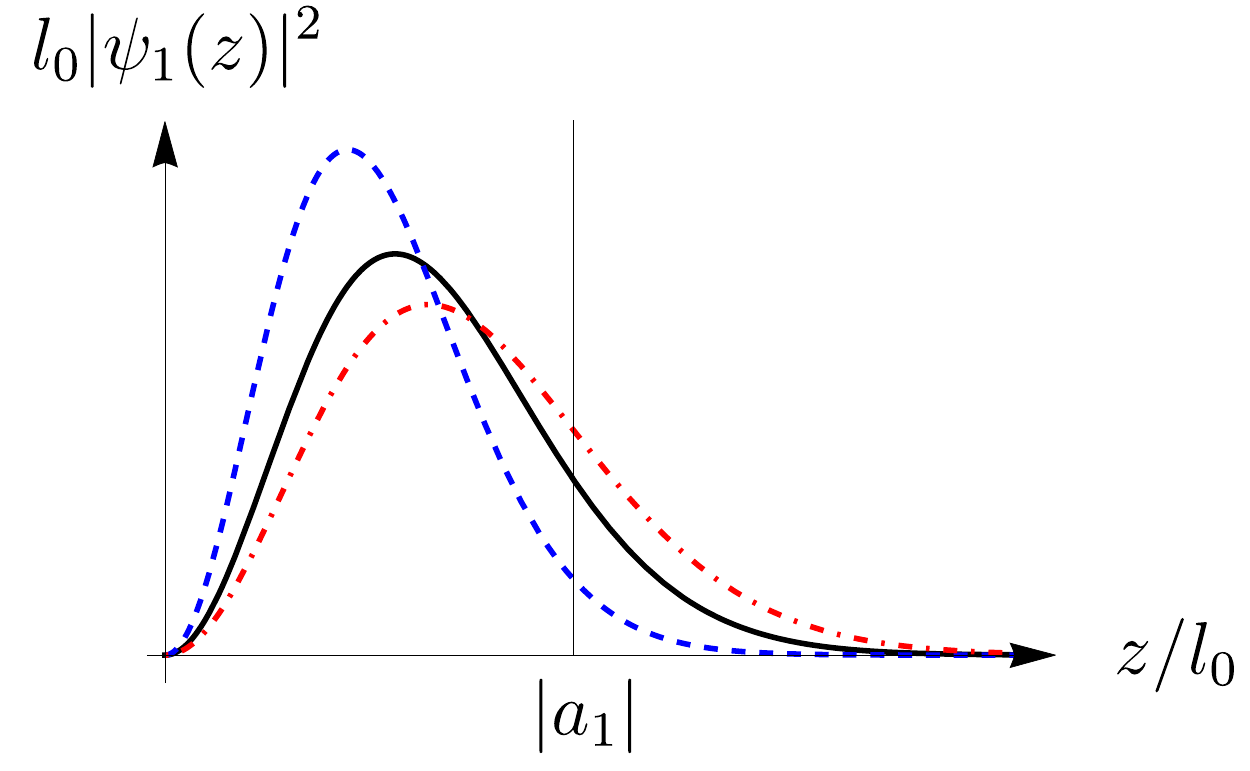} \qquad
\includegraphics[scale=0.5]{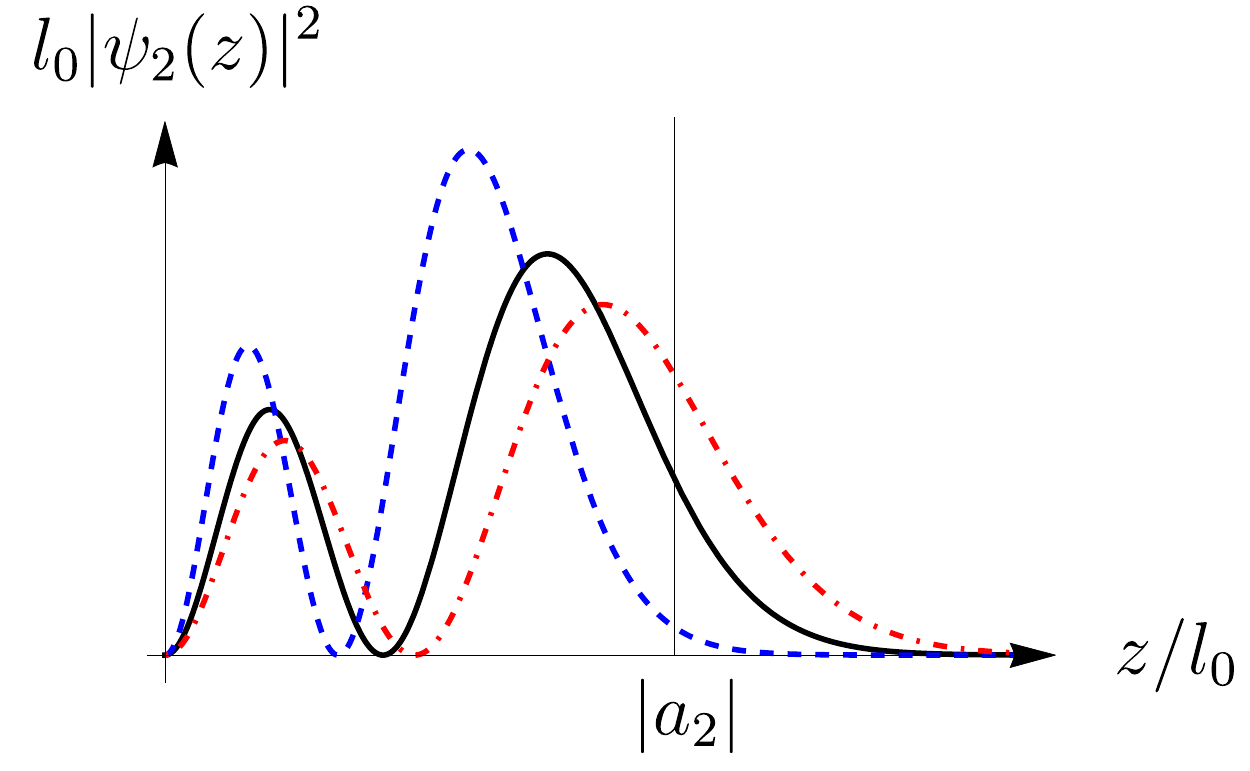}\\[8pt]
\includegraphics[scale=0.5]{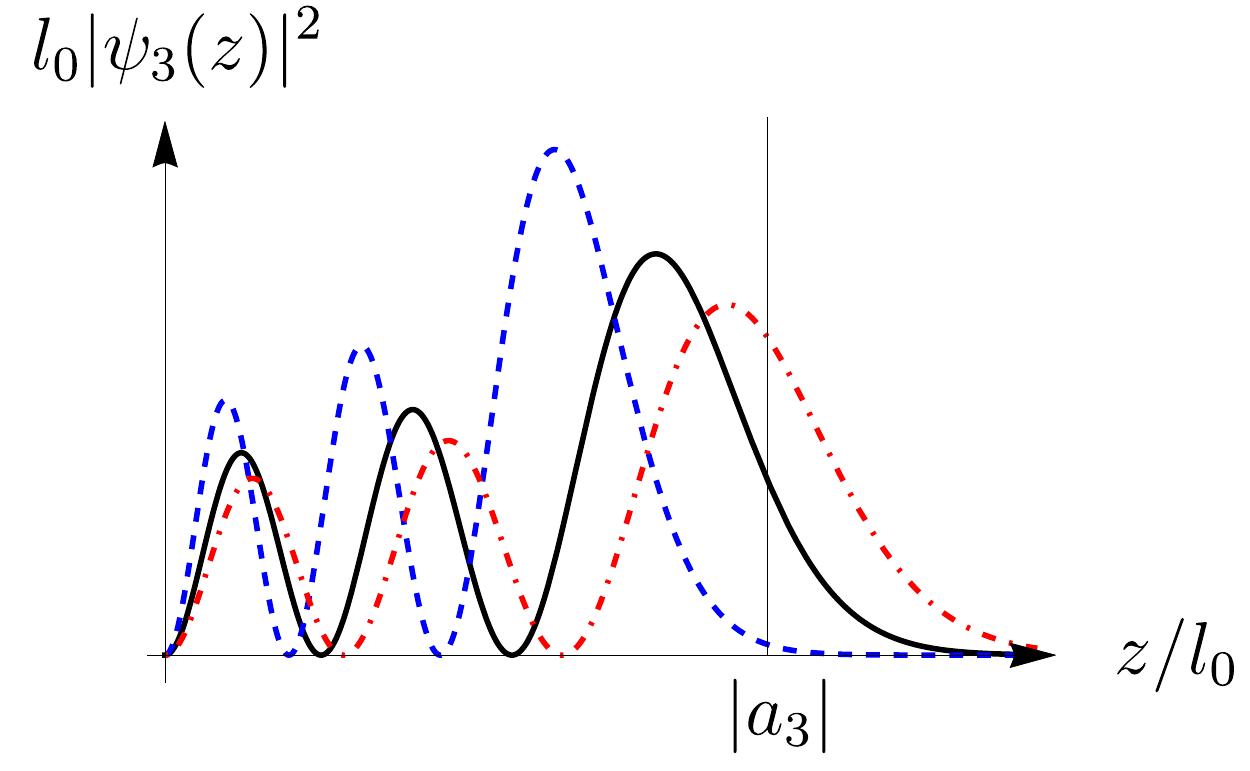} \qquad
\includegraphics[scale=0.5]{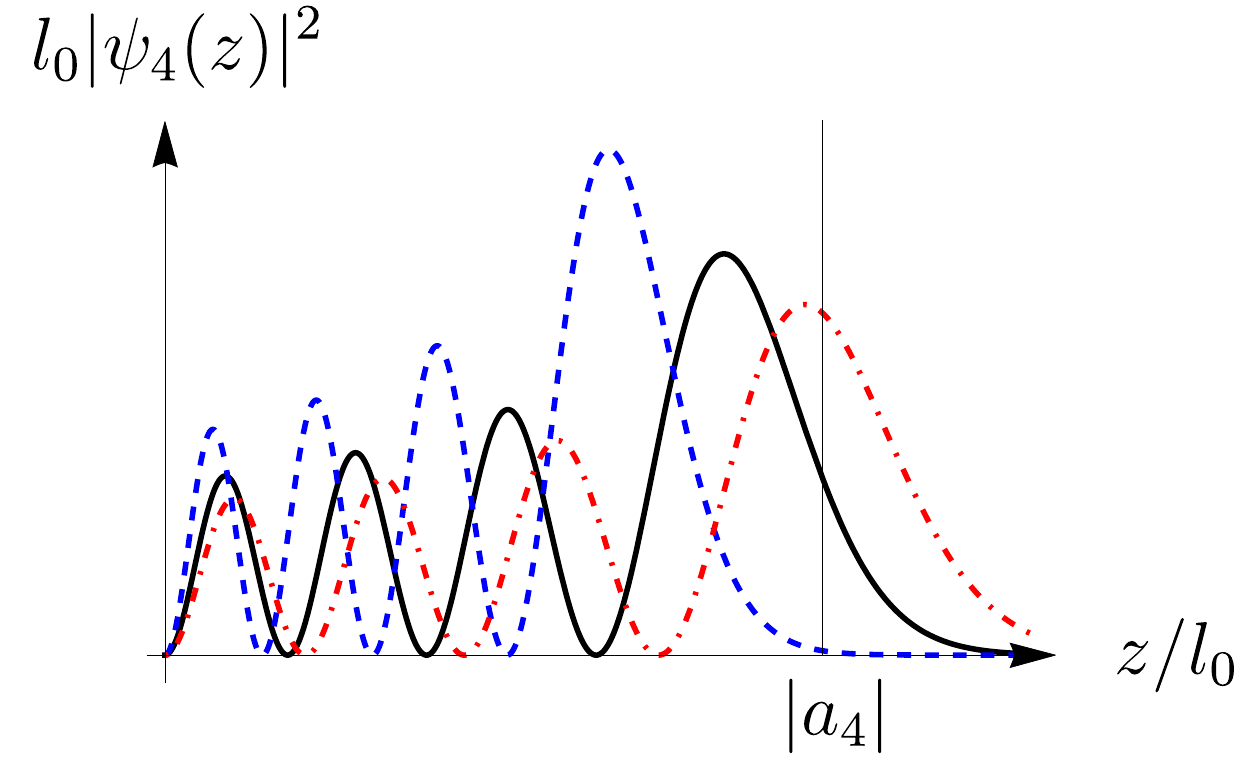}
\caption{Squared moduli of the neutron wave functions $\vert \psi _{n}(z) \vert ^{2}$ are shown as a function of the  height $z$ for the four lowest quantum states; they correspond to the probabilities of observing neutrons. The black continuous line corresponds to the Lorentz-symmetric case ($\xi = 0$),  the blue dashed line displays the LV case $\xi > 0$, and the red dot-dashed line shows the LV case $\xi < 0$.} \label{ProbDensityPlots}
\end{figure}

In Figure \ref{ProbDensityPlots} we show the probability density $\vert \psi _{n}(z) \vert ^{2}$ for the four lowest quantum states $n=1,2,3,4$ of ultracold neutrons. The black continuous line corresponds to the Lorentz-symmetric case ($\xi = 0$),  the blue dashed line displays the LV case $\xi > 0$, and the red dot-dashed line shows the LV case $\xi < 0$.  On the one hand, we observe that the LV parameters $(\mathfrak {p},\xi)$ do not affect the number of nodes in the probability density.  The exponential factor in Eq. (\ref{Wave-Function-Fin}) is never zero, and hence the number of nodes is fully controlled by the Airy function. On the other hand, we can see that the position of the nodes are shifted downwards for $\xi < 0$ and upwards for $\xi > 0$. We can understand this from Eq. (\ref{EnergySpectrum}), since it can be interpreted as the Lorentz-symmetric energy spectrum with an effective gravitational acceleration of the form $g _{\mbox{\scriptsize eff}} = g (1-\xi ) ^{1/3}$.  This result is far from obvious, since the gravitational potential in the Hamiltonian is not directly affected by the  parameters of violation of Lorentz invariance. Clearly, $g _{\mbox{\scriptsize eff}} > g$ for $\xi <0$, thus implying that neutrons are pulled down stronger than in the standard case. This explains why the nodes are shifted down in this case. The opposite case can be understood in a similar manner.  For $\xi >0$ we get $g _{\mbox{\scriptsize eff}} < g$,  thus suggesting that neutrons fall slower than in the standard Lorentz-symmetric case, and hence the nodes are shifted upwards.

Using the quantum states of UCNs (\ref{Wave-Function-Fin}) in the presence of Lorentz-violation, one can immediately compute the LV effects on physically relevant average values. In particular, the expectation value of any function $f(z)$ is $\mathfrak {p}$-independent. For example,  the average value of the $k$-th power of the position, with $k \geq 0$, can be related with the Lorentz-symmetric result as follows
\begin{align}
\braket{z ^{k}}_{n} = \int _{0} ^{\infty} z ^{k} \vert \psi _{n} (z) \vert ^{2} dz = (1 - \xi ) ^{k/3} \braket{z ^{k}}_{n} \!\!\!\! {}^{(0)} ,
\end{align}
where $\braket{z ^{k}}_{n} \!\!\!\! {}^{(0)} $ is the expectation value at $\xi=0$. Recall that $\braket{z}_{n} \!\!\!\! {}^{(0)} = \frac{2}{3} h _{n}$, and $\braket{z ^{2}}_{n} \!\!\!\! {}^{(0)} = \frac{8}{15} h _{n} ^{2}$, where $h _{n} = - a _{n} l _{0}$. The average value of the momentum operator is subtler due to the exponential factor appearing in the quantum states (\ref{Wave-Function-Fin}).  For instance, the average of momentum is
\begin{align}
\braket{\hat{p} _{z}}_{n} = \int _{0} ^{\infty} \psi ^{\ast} (z)  \hat{p} _{z} \psi (z)dz \ = \ \int _{0} ^{\infty} \phi (z)  \left( \mathfrak{p} + \hat{p} _{z} \right) \phi (z)dz \ = \ \mathfrak{p}\ ,
\end{align}
which, in general, is nonzero.  In the standard case $\mathfrak{p}=\xi=0$, the average momentum vanishes identically due to the parity symmetry.  However, in the presence of Lorentz violation, the Hamiltonian (\ref{HamLV}) breaks the parity explicitly. In a similar fashion we evaluate the expectation value of the square of momentum:
\begin{align}
\braket{\hat{p} _{z}^{2}}_{n} = \int _{0} ^{\infty} \psi ^{\ast} (z)  \hat{p} _{z}^{2} \psi (z)dz = \int _{0} ^{\infty} \phi (z)  \left( \mathfrak{p} + \hat{p} _{z} \right)^{2} \phi (z)dz = \mathfrak{p} ^{2} + (1- \xi ) ^{-2/3} \braket{p _{z} ^{2}}_{n} \!\!\!\! {}^{(0)} ,
\end{align}
here $\braket{p _{z} ^{2}}_{n} \!\!\!\! {}^{(0)} = \frac{1}{3} m ^{2}gh _{n}$ is the expectation value at $\mathfrak{p}=\xi=0$. These expectation values can be connected by the virial theorem as usual.

\section{Comparison with experimental data} \label{ExpData}

\subsection{The GRANIT experiment}

The quantization of neutron states bouncing on a mirror, due to the Earth's gravitational field, have been investigated in the GRANIT experiments. Such quantization have been confirmed, with high statistical accuracy and methodological reliability, by measuring the characteristic sizes of the neutron wave functions in the first and second quantum states \cite{Nesvizhevsky_2005}.  The measured values 
\begin{align}
h _{1} ^{\textrm{exp}} = ( 12.2 \pm 1.8 _{\textrm{sys}} \pm 0.7_{\textrm{stat}} ) \mu \mbox{m} ,  \qquad h _{2} ^{\textrm{exp}} = (21.6 \pm 2.2 _{\textrm{sys}} \pm 0.7_{\textrm{stat}} ) \mu \mbox{m}
\end{align}
agree with the theoretical values $h _{1} = 13.7 \,  \mu$m and $h _{2} = 24.0 \,  \mu$m,  calculated as $h _{n} = E _{n} ^{(0)} / mg = - a _{n} l _{0}$, where $E _{n} ^{(0)}=-mgl _{g} a _{n}$ are the unperturbed energy levels (i.e. they are obtained from the energy spectrum (\ref{EnergySpectrum}) with $\xi = 0$) \cite{Nesvizhevsky_2002}. So, the relative experimental uncertainties should be attributed to the measurement of the energy.  As the experimental results supported the quantum mechanical expectation without additional interactions other than gravity, the good agreement between theory and experiment can be turned into constraints onto unconventional physics.  For instance, interesting constraints on fundamental short-range forces \cite{Nesvizhevsky_2008, Baeler_2009, Antoniadis_2011}, on axion-like interactions \cite{Baeler_2007},  on the electric neutron charge \cite{Borisov_1988} and on the fundamental length scale in polymer quantum mechanics \cite{Martin_2015} can be obtained using quantum gravitational states of UCNs. They have also been used, within the framework of the SME, to set upper bounds to the LV parameters of the fermion sector of the SME \cite{Martin_2018, Xiao_2020, Ivanov_2019} and those characterizing the gravitational sector as well \cite{Escobar_2019, Ivanov_2021}.  In this Section we follow these ideas and compare the theoretical results of the previous Section \ref{QGS-UCNs} with the reported data in the GRANIT experiment to set an upper bound for the fluctuations $\delta h _{\mu \nu}$.  To this end, we use the maximal experimental error in the energy spectrum $\vert \Delta E _{n} ^{\textrm{exp}} \vert $ to constraint deviations of the exact energy spectra $E_{n}$, given by Eq. (\ref{EnergySpectrum}), from the well-measured Lorentz symmetric case $E_{n} ^{(0)}$, i.e.
\begin{align}
\vert E_{n} - E_{n} ^{(0)} \vert < \vert \Delta E _{n} ^{\textrm{exp}} \vert .  \label{Rel1a}
\end{align}
The current sensitivity of the GRANIT experiment on the energy levels of the ground and first excited quantum gravitational states of UCNs,  $\vert \Delta E _{1} ^{\textrm{exp}} \vert = 0.102$ peV and $\vert \Delta E _{2} ^{\textrm{exp}} \vert = 0.051$ peV \cite{Nesvizhevsky_2005}, respectively,  suggest that deviation from Lorentz invariance has to be small. Consequently,  assuming $\xi  \ll 1$, we can power expand the exact energy levels (\ref{EnergySpectrum}). Up to leading order we obtain $\vert E_{n} - E_{n} ^{(0)} \vert \approx \frac{1}{3}  \vert \xi \vert  E_{n} ^{(0)}$, and upon substitution of this result into Eq.  (\ref{Rel1a}) we get
\begin{align}
\vert \delta h _{00} + \delta h _{zz} \vert < 3\,  \frac{ \vert \Delta E _{n} ^{\textrm{exp}} \vert }{ E_{n} ^{(0)}  }  .   \label{Rela1ab}
\end{align}
Using the experimental data for the energy levels of the two lowest quantum gravitational states of UCNs, together with the theoretical values $E_{1} ^{(0)}=1.407$peV and $E_{2} ^{(0)}=2.461$peV, we impose an upper bound on the parameters of violation of Lorentz invariance, namely $\vert \delta h _{00} + \delta h _{zz} \vert < 10 ^{-2}$.   

\subsection{The $q$Bounce experiment}

An interesting possibility to set a slightly better bound on the LV coefficients is from the analysis of experimental data on transition frequencies between quantum gravitational states of unpolarized UCNs, reported in Ref.  \cite{Cronenberg_2018} by the $q$Bounce collaboration.   The transition frequency $\nu _{m \to n}$ of the transition between two gravitational states of polarized UCNs $\ket{m} \, \to \, \ket{n}$ is defined by $\nu _{m \to n} = (E_{n}-E_{m})/ 2 \pi$, where $E_{n}$ and $E_{m}$ are observable binding energies of UCNs in the $\ket{n}$ and $\ket{m}$ quantum gravitational states, respectively.  Contributions to the transition frequencies due to violation of Lorentz  invariance are defined by corrections to the binding energies of unpolarized UCNs. In the present case, since $E_{n}\approx E_{n} ^{(0)} (1 - \xi / 3)$, for $\xi \ll 1$, the transition frequency in the presence of Lorentz violation $\nu _{m \to n}$ can be related to the unperturbed transition frequency $\nu _{m \to n} ^{(0)} = (E_{n} ^{(0)} -E_{m} ^{(0)})/ 2 \pi$ by
\begin{align}
\nu _{m \to n} = (1 - \xi / 3) \, \nu _{m \to n} ^{(0)}  . 
\end{align}
The theoretical values
\begin{align}
\nu _{3 \to 1} ^{(0)} = 0.3047 \, \mbox{peV} ,  \qquad \nu _{4 \to 1} ^{(0)} =  0.4259 \, \mbox{peV} ,
\end{align} 
are in good agreement with the experimental results of the $q$Bounce experiment \cite{Cronenberg_2018}:
\begin{align}
\nu _{3 \to 1} ^{\textrm{exp}} = 0.3059(8) \, \mbox{peV} ,  \qquad \nu _{4 \to 1} ^{\textrm{exp}} =  0.4277(12) \, \mbox{peV} .
\end{align} 
The experimental values of these transition frequencies are measured with relative uncertainties $2.6 \times 10 ^{-3}$ and $2.8 \times 10 ^{-3}$, respectively.  The relative experimental uncertainties define the current sensitivity of the $q$Bounce experiments for the non-spin-flip transition frequencies, which in this case is $\Delta \nu < 0.3183 \times 10 ^{-15}$ eV.  Upon comparison of the theoretical and experimental values for the transition frequencies we get an upper bound for the Lorentz-violating coefficients:
\begin{align}
\vert \delta h _{00} + \delta h _{zz} \vert < 3 \frac{\vert \Delta \nu \vert }{ \nu _{m \to n} ^{(0)} } . 
\end{align}
Using the theoretical and experimental values for the transitions $\ket{1} \, \to \, \ket{3}$ and $\ket{1} \, \to \, \ket{4}$ we get the constraints $\vert \delta h _{00} + \delta h _{zz} \vert < 2 \times 10 ^{-3}$ and $\vert \delta h _{00} + \delta h _{zz} \vert < 3 \times 10 ^{-3}$, respectively. These are one order of magnitude better than the bound imposed by the GRANIT experiment. As asserted in Ref.  \cite{Abele_2010}, the $q$Bounce experiments will improve the experimental sensitivity in the nearest future up to $10 ^{-17}$ eV, and hence it should allow to improve by two order of magnitude the upper bound for the combination of parameters $ \delta h _{00} + \delta h _{zz} $.  Improvement of such bounds is possible,  as we shall discuss later.

\subsection{Bound of violation of Lorentz invariance in the canonical Sun-centered frame}

For comparative purposes, a standard inertial frame is used to report the Lorentz violation coefficients. The canonical frame chosen is the Sun-centered celestial equatorial frame \cite{Kostelecky_2002}. Taking in the laboratory the $x$ axis pointing to the south, the $y$ axis to east, and the $z$ axis vertically upwards and with the reasonable approximation that the Earth's orbit is circular, the spatial transformation between the laboratory frame and the Sun-centered frame is given by
\begin{align}
R_{jJ}=\begin{pmatrix}
\cos\chi \cos\omega_{\oplus} T_{\oplus} & \cos\chi \sin\omega_{\oplus} T_{\oplus} & -\sin\chi \\
-\sin\omega_{\oplus} T_{\oplus} & \cos\omega_{\oplus} T_{\oplus}  & 0\\
\sin\chi \cos\omega_{\oplus} T_{\oplus} & \sin\chi \sin\omega_{\oplus} T_{\oplus} & \cos{\chi}
\end{pmatrix},  \label{RRR}
\end{align}
being $j=x,y,z$ an index in the laboratory frame and $J=X,Y,Z$ an index in the Sun-centered frame. The quantities $\omega_{\oplus}\approx2\pi/(23\textrm{ h } 56 \textrm{ min})$, $\chi$ and $T_{\oplus}$ correspond to the Earth's sideral frequency, the colatitude of the laboratory and the local sideral time, which is properly chosen for each experiment. The combination of the coefficients, in the Sun-centered frame, that we obtain in the present work has the following form
\begin{align}
\vert \delta h _{00} - \delta h _{zz} \vert =& \bigg| \delta h _{TT} - \left[ \frac{1}{2}( \sin ^{2} \chi( \delta h _{XX} + \delta h _{YY}) + 2 \cos ^{2} \chi \delta h _{ZZ}) \right.  \\ \nonumber & \left. + 2 \sin \chi \cos \chi ( \delta h _{XZ} \cos \omega _{\oplus} T_{\oplus} + \delta h _{YZ} \sin \omega _{\oplus} T_{\oplus} ) + \frac{1}{2} \sin ^{2} \chi ( \delta h _{XX} - \delta h _{YY} ) \cos 2 \omega _{\oplus} T _{\oplus} \right] \bigg| .
\end{align}
The origin of the time $t$ should be defined for a given laboratory.  An appropriate choice is to match $t$ with the local sideral time $T_{\oplus}$, at the moment when the axis $y$, in the laboratory, and $Y$, in the Sun-centered frame, coincide. The relation between the local sideral time $T_{\oplus}$ and the celestial equatorial time $T$ is given by \cite{Ding_2016}
\begin{align}
T _{\oplus} = T - T _{0} ,  \quad\quad T _{0} = \frac{66.25^{\circ} - \phi }{360^{\circ}}(23.934 \textrm{hr}),
\end{align}
where $\phi$ is a longitude of the laboratory measured in degrees. Employing the colatitude of the Grenoble's laboratory $\chi=44.83^{\circ}$ we can establish the upper bound
\begin{align}
\vert 0.502 \delta h _{XX} + \delta h _{YY} + 0.497 \delta h _{ZZ} - 0.999 \delta h _{XZ} \vert < 10 ^{-2}, \label{BoundSun-Centered}
\end{align}
where we have employed the traceless condition $\delta h _{TT} = \delta h _{XX} + \delta h _{YY} + \delta h _{ZZ}$. In principle, by orientating the experimental device at different positions and performing the experiment at different times we can obtain different combinations of the Lorentz-violating coefficients; however, the bound will be the same order of magnitude, $\sim 10 ^{-2}$.

\section{COW test of Lorentz invariance violation} \label{COW-sect}

\begin{figure}
\includegraphics[scale=0.4]{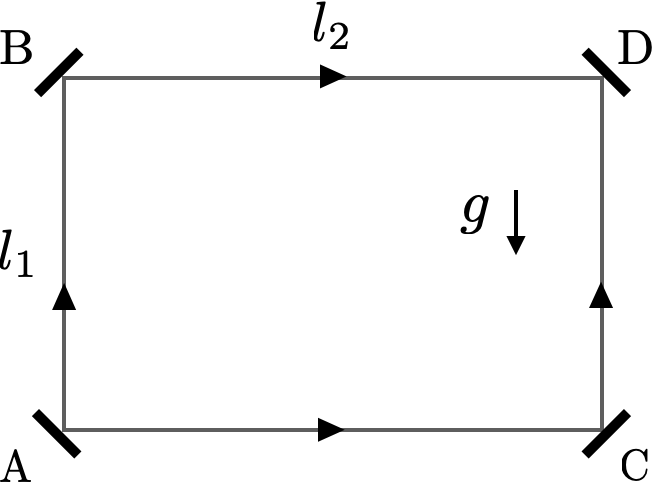}
\caption{Schematics of the COW experiment.} \label{COW}
\end{figure}

In this Section we suggest that the neutron interferometric experiment of Collela, Overhauser and Werner \cite{COW_1975} (which we refer to as COW experiment) is sensitive to violations of Lorentz invariance, and in this case can set up a conservative upper bound on the LV coefficients. The COW experiment was designed to demonstrate the validity of the weak equivalence principle using the gravitationally induced quantum-mechanical phase shift in the interference between coherently split and separated neutron de Broglie waves. Our description of the COW experiment follows the Fig. \ref{COW}. A monochromatic neutron beam is sent, from A to D, by two different paths on a plane, ABD and ACD, as shown in Fig. \ref{COW}.  Clearly, the horizontal paths AC and BD are at different gravitational potentials. The resulting interference pattern at detector in D is completely attributed to the gravity-induced difference in the wavelength of the two horizontal paths. If we consider that no anomalous effect has been observed in the experiment, any new mechanism predicting deviations from the phase shift must be smaller that one fringe. This idea has been used to constraint possible deviations from Newtonian gravity \cite{Bertolami_1986}, noncommutative structure of spacetime \cite{Saha_2014}, and generalized uncertainty principles \cite{Xiang_2018, Farahani_2020}.

In a COW-type experimental setting, the gravitational potential is much smaller than the total energy of the neutrons and we can calculate the gravity-induced phase shift from (\ref{HamLV}) by the semiclassical prescription of matter-wave interferometry. Our starting point is the corresponding classical Hamiltonian: $h(z,p_{z}) = \frac{ p _{z} ^{2}}{2\, m} (1 - \xi )+ mgz $, which we read from Eq. (\ref{Hhz}). The general solution to the equations of motion
\begin{align}
\dot{z} = \frac{p}{m} (1 - \xi )  , \qquad \dot{p} = - m \,g ,  
\end{align}
can be described by
\begin{align}
z (t) = a + b \, t - \frac{1}{2} g \, t^{2} (1 - \xi ),  \label{ClassSol}
\end{align}
where $a$ and $b$ are arbitrary constants.  Imposing the end-point conditions $z _{0} = z (t _{0})$ and $z _{1} = z (t _{1})$ we get
\begin{align}
a = \frac{x_{0} t _{1} - x _{1} t _{0}}{t _{1} - t _{0}} - \frac{1}{2} g  t _{0} t _{1} (1 - \xi )\  , \qquad b = \frac{x _{1} - x _{0}}{t _{1} - t _{0}} + \frac{1}{2} g ( t _{1} + t _{0} ) (1 - \xi ) . \label{Coef-ClassSol}
\end{align}
The action associated with any test path $z(t)$ is given by $S[z(t)] = \int _{t_{0}} ^{t_{1}} L \left( z(t),\dot{z}(t) \right) dt $, where $L(z,\dot{z})$ is the Lagrangian. Taking the solution described by Eq. (\ref{ClassSol}) with the coefficients (\ref{Coef-ClassSol}) we find
\begin{align}
S(z_{1},t_{1};z_{0},t_{0}) = \frac{m}{2 (t _{1} - t _{0} ) (1- \xi ) }   {(z _{1} - z _{0})}^{2}  -  \frac{1}{2} m g (t _{1} - t _{0} ) (z _{1} + z _{0}) (1- \xi )  -  \frac{m g^{2} (t _{1} - t _{0} )^{3} }{24}  (1- \xi )( 1 - 2 \xi ) .  \label{ActionFunc}
\end{align}
This is the action functional associated with the dynamical path linking those spacetime points $(z_{1},t_{1})$ and $(z_{0},t_{0})$. Now let us consider two interfering neutron beams: ABD and ACD, as shown in Fig. \ref{COW}.  For simplicity, the plane ABCD is set to be vertical to the horizontal plane.  We now investigate the effects of Lorentz invariance violation upon the phase difference measured by the COW experiment.  Using the action functional we now compute the phase difference.  The phase acquired along the path ABD is
\begin{align}
S_{\mbox{\scriptsize{ABD }}}= \frac{m}{2} \left[ \frac{l _{1}^{2}}{ t _{1} (1- \xi )} + \frac{l _{2} ^{2}}{t - t _{1}} \right] -  mg l _{1} \left( t - \tfrac{1}{2} t _{1} \right)(1- \xi ) -  \frac{m g^{2} t _{1} ^{3} }{24}  (1- \xi )( 1 - 2 \xi )
\end{align}
while for the path ACD
\begin{align}
S_{\mbox{\scriptsize{ACD }}}=  \frac{m}{2} \left[ \frac{l _{1}  ^{2}}{ t _{1} (1- \xi )} + \frac{l _{2} ^{2}}{t - t _{1}} \right] - \frac{1}{2} mg l _{1} t _{1} (1- \xi ) -  \frac{m g^{2} t _{1} ^{3} }{24}  (1- \xi )( 1 - 2 \xi ) ,
\end{align}
$t_{1}$ is the vertical travel time and $t-t_{1}$ is the horizontal travel time. It is clear that between the two amplitudes there is a phase difference
\begin{align}
\Delta \phi = \frac{1}{\hbar} (S_{\mbox{\scriptsize{ABD }}} - S_{\mbox{\scriptsize{ACD }}})=- 2 \pi \lambda (m/ \hbar ) ^{2} g (1- \xi ) l_{1}l_{2} ,   \label{PhaseShift}
\end{align}
where we have replaced the horizontal travel time in terms of the de Broglie wave-length $\lambda$, namely, $t-t_{1}=l_{2} m \lambda / h$. 

The first experimental demonstration of the gravitationally induced phase shift was reported by COW in 1975 \cite{COW_1975}. The agreement with theory was 90\%.  Later, more detailed investigations were carried out in which simultaneous effects of gravity and inertia on the motion of neutrons were considered \cite{COW_1980, Bonse_1983, Bonse_1984, Werner_1988}, providing convincing high-quality data deviating from theory by about 1\%.  In a more recent experiment, performed with a pair of almost harmonic wavelengths, the obtained values and the theoretical prediction showed a discrepancy only about 0.1\% \cite{Littrell_1997}. Therefore, any hypothetical deviation from the gravitationally induced phase shift must be contained within the experimental error.  We now use the maximal experimental error in the energy shift $ \vert \Delta \phi ^{\mbox{\scriptsize exp}} \vert$ to constraint deviations of the exact result, given by Eq. (\ref{PhaseShift}), from the well-measured Lorentz symmetric case $\Delta \phi ^{(0)} = - 2 \pi \lambda (m/ \hbar ) ^{2} g l_{1}l_{2}$, i.e.
\begin{align}
\vert \Delta \phi -  \Delta \phi ^{(0)} \vert < \vert \Delta \phi ^{\mbox{\scriptsize exp}} \vert . 
\end{align}
Using the experimental data we get the conservative constraint $\vert  \delta h _{00}+\delta h _{zz} \vert < 10^{-3}$.  Therefore, the COW experiment yields a bound that is of the same order of magnitude than the one obtained with the $q$Bounce experiment, and better than the one obtained with the GRANIT experiment. In the Sun-centered frame, the bound reads as in Eq. (\ref{BoundSun-Centered}).

\section{Summary and conclusions}
\label{Conclu1}

This work can be divided in two parts. In the first part (Section \ref{QGS-UCNs}) we investigate, within the scalar sector of the Standard-Model Extension, the quantum gravitational states of ultracold neutrons in a Lorentz violating background. We parametrize the Lorentz-violation in terms of a symmetric tensor $h_{\mu \nu}$ that represents a constant background. In light of the recent experimental advances with UCNs in the Earth's gravitational field,  we frame our model according to the laboratory conditions under which GRANIT and $q$Bounce experiments are performed (i.e. nonrelativistic velocities and semiclassical behaviour in the perpendicular direction to the gravitational acceleration).  We solve the corresponding Schr\"{o}dinger equation in an analytical fashion and determine the energy eigenstates and eigenvalues. We show that the effects of Lorentz invariance violation can be interpreted in terms of an effective gravitational acceleration $g_{\mbox{\scriptsize eff}}=g(1-\xi)^{1/3}$, with $\xi = \delta h _{00}+\delta h _{zz}$, which can be either larger or smaller than $g$, depending on the sign of $\xi$.  As evinced by the oscillations in the probability distribution plots, in the former case UCNs fall down more slowly, whilst in the latter case they fall down more fastly.  This interpretation is far from obvious if we directly see the effective Hamiltonian. Also, as expected, the energy spectrum in the presence of Lorentz violation is shifted with respect to the standard energy levels ($\xi = 0$).

Using the experimental sensitivity of GRANIT and $q$Bounce experiments we make some estimates of upper bounds of the parameter $\xi$ of Lorentz invariance violation in the scalar sector of the SME.  On the one hand, the GRANIT experiment specializes in the measurement of the height of the lowest bound states of UCNs, confirming in this way the predictions of quantum mechanics. The $q$Bounce experiment, on the other hand,  concentrates on the measurement of transition frequencies between gravitational quantum states. Since the data shows good agreement with theory, one can constraints deviations from quantum physics due to eventual new physical mechanisms. In recent papers \cite{Martin_2018, Escobar_2019},  two of us proposed that Lorentz and CPT invariance can be tested with UCNs, both in the fermion and gravitational sectors of the SME. The appropriate generalization to include the spin came fastly in Refs.  \cite{Ivanov_2019, Xiao_2020, Ivanov_2021}. Along this light, here we propose that the scalar sector of the SME can also be tested by using unpolarized UCNs. A salient feature in this case is that Schr\"{o}dinger equation can be solved analytically, unlike the previously cited works, where energy shifts are computed within a perturbative scheme.  We have also verified that our results are consistent with that of perturbation theory for small departures from Lorentz symmetry. Using the current sensitivity $\Delta E < 2 \times 10 ^{-15}$eV of the GRANIT and $q$Bounce experiments, we set an upper bound for a particular combination of the Lorentz violating coefficients, namely: $\vert  \delta h _{00}+\delta h _{zz} \vert < 10^{-3}$, which we expressed also in the canonical Sun-centered frame. This result improves, by one order of magnitude, the bound imposed by two of us with the use of Casimir physics on a different combination of these coefficients \cite{Escobar_2020}. The constraints we report here can be substantially improved in the nearest future by the enhancement of the sensitivity of the $q$Bounce experiment up to $\Delta E < 10 ^{-17}$eV \cite{Abele_2010}. 

In the second part of this paper (Section \ref{COW-sect}) we discuss the possibility for testing Lorentz invariance violation with neutron interferometry, namely, the COW experiment. To this end, using the semiclassical prescription of matter-wave interferometry, we evaluate the gravity-induced phase shift, and upon comparison with the current experimental sensitivity, we find an upper bound similar to that obtained with the $q$Bounce experiment. To the best of our knowledge, COW-type experiments has been used to test deviations from Newtonian gravity and quantum mechanics, but this is the first time it is used to test Lorentz invariance violation. In this case, we test the validity of such symmetry by using a nonpolarized beam of neutrons (since this theory is scalar), however,  COW-type experiments can also be used within the fermion sector of the SME, where neutron spin would play a fundamental role. It can be used also to test the gravitational sector of the SME. In any case, conservative bounds on the parameters will be obtained. We leave these ideas for future works.

\acknowledgements

A. M.-R. acknowledges support from DGAPA-UNAM Project No. IA102722 and by Project CONACyT (M\'{e}xico) No. 428214.  C. A. E. is supported from Project PAPIIT No. IN109321.

\bibliography{Bib.bib}

\begin{thebibliography}{63}%
\makeatletter
\providecommand \@ifxundefined [1]{%
 \@ifx{#1\undefined}
}%
\providecommand \@ifnum [1]{%
 \ifnum #1\expandafter \@firstoftwo
 \else \expandafter \@secondoftwo
 \fi
}%
\providecommand \@ifx [1]{%
 \ifx #1\expandafter \@firstoftwo
 \else \expandafter \@secondoftwo
 \fi
}%
\providecommand \natexlab [1]{#1}%
\providecommand \enquote  [1]{``#1''}%
\providecommand \bibnamefont  [1]{#1}%
\providecommand \bibfnamefont [1]{#1}%
\providecommand \citenamefont [1]{#1}%
\providecommand \href@noop [0]{\@secondoftwo}%
\providecommand \href [0]{\begingroup \@sanitize@url \@href}%
\providecommand \@href[1]{\@@startlink{#1}\@@href}%
\providecommand \@@href[1]{\endgroup#1\@@endlink}%
\providecommand \@sanitize@url [0]{\catcode `\\12\catcode `\$12\catcode
  `\&12\catcode `\#12\catcode `\^12\catcode `\_12\catcode `\%12\relax}%
\providecommand \@@startlink[1]{}%
\providecommand \@@endlink[0]{}%
\providecommand \url  [0]{\begingroup\@sanitize@url \@url }%
\providecommand \@url [1]{\endgroup\@href {#1}{\urlprefix }}%
\providecommand \urlprefix  [0]{URL }%
\providecommand \Eprint [0]{\href }%
\providecommand \doibase [0]{https://doi.org/}%
\providecommand \selectlanguage [0]{\@gobble}%
\providecommand \bibinfo  [0]{\@secondoftwo}%
\providecommand \bibfield  [0]{\@secondoftwo}%
\providecommand \translation [1]{[#1]}%
\providecommand \BibitemOpen [0]{}%
\providecommand \bibitemStop [0]{}%
\providecommand \bibitemNoStop [0]{.\EOS\space}%
\providecommand \EOS [0]{\spacefactor3000\relax}%
\providecommand \BibitemShut  [1]{\csname bibitem#1\endcsname}%
\let\auto@bib@innerbib\@empty
\bibitem [{\citenamefont {Colladay}\ and\ \citenamefont
  {Kosteleck\'y}(1997)}]{Colladay_Kostelecky_1997}%
  \BibitemOpen
  \bibfield  {author} {\bibinfo {author} {\bibfnamefont {D.}~\bibnamefont
  {Colladay}}\ and\ \bibinfo {author} {\bibfnamefont {V.~A.}\ \bibnamefont
  {Kosteleck\'y}},\ }\bibfield  {title} {\bibinfo {title} {C{PT} violation and
  the standard model},\ }\href {https://doi.org/10.1103/PhysRevD.55.6760}
  {\bibfield  {journal} {\bibinfo  {journal} {Phys. Rev. D}\ }\textbf {\bibinfo
  {volume} {55}},\ \bibinfo {pages} {6760} (\bibinfo {year}
  {1997})}\BibitemShut {NoStop}%
\bibitem [{\citenamefont {Colladay}\ and\ \citenamefont
  {Kosteleck\'y}(1998)}]{Colladay_Kostelecky_1998}%
  \BibitemOpen
  \bibfield  {author} {\bibinfo {author} {\bibfnamefont {D.}~\bibnamefont
  {Colladay}}\ and\ \bibinfo {author} {\bibfnamefont {V.~A.}\ \bibnamefont
  {Kosteleck\'y}},\ }\bibfield  {title} {\bibinfo {title} {Lorentz-violating
  extension of the standard model},\ }\href
  {https://doi.org/10.1103/PhysRevD.58.116002} {\bibfield  {journal} {\bibinfo
  {journal} {Phys. Rev. D}\ }\textbf {\bibinfo {volume} {58}},\ \bibinfo
  {pages} {116002} (\bibinfo {year} {1998})}\BibitemShut {NoStop}%
\bibitem [{\citenamefont {Bailey}\ and\ \citenamefont
  {Kosteleck\'y}(2006)}]{Bailey_Kostelecky_2006}%
  \BibitemOpen
  \bibfield  {author} {\bibinfo {author} {\bibfnamefont {Q.~G.}\ \bibnamefont
  {Bailey}}\ and\ \bibinfo {author} {\bibfnamefont {V.~A.}\ \bibnamefont
  {Kosteleck\'y}},\ }\bibfield  {title} {\bibinfo {title} {Signals for
  {L}orentz violation in post-{N}ewtonian gravity},\ }\href
  {https://doi.org/10.1103/PhysRevD.74.045001} {\bibfield  {journal} {\bibinfo
  {journal} {Phys. Rev. D}\ }\textbf {\bibinfo {volume} {74}},\ \bibinfo
  {pages} {045001} (\bibinfo {year} {2006})}\BibitemShut {NoStop}%
\bibitem [{\citenamefont {Kosteleck\'y}\ and\ \citenamefont
  {Tasson}(2011)}]{Kostelecky_Tasson_2011}%
  \BibitemOpen
  \bibfield  {author} {\bibinfo {author} {\bibfnamefont {V.~A.}\ \bibnamefont
  {Kosteleck\'y}}\ and\ \bibinfo {author} {\bibfnamefont {J.~D.}\ \bibnamefont
  {Tasson}},\ }\bibfield  {title} {\bibinfo {title} {Matter-gravity couplings
  and {L}orentz violation},\ }\href
  {https://doi.org/10.1103/PhysRevD.83.016013} {\bibfield  {journal} {\bibinfo
  {journal} {Phys. Rev. D}\ }\textbf {\bibinfo {volume} {83}},\ \bibinfo
  {pages} {016013} (\bibinfo {year} {2011})}\BibitemShut {NoStop}%
\bibitem [{\citenamefont {Kosteleck\'y}\ and\ \citenamefont
  {Samuel}(1989)}]{Kostelecky_1989}%
  \BibitemOpen
  \bibfield  {author} {\bibinfo {author} {\bibfnamefont {V.~A.}\ \bibnamefont
  {Kosteleck\'y}}\ and\ \bibinfo {author} {\bibfnamefont {S.}~\bibnamefont
  {Samuel}},\ }\bibfield  {title} {\bibinfo {title} {Spontaneous breaking of
  {L}orentz symmetry in string theory},\ }\href
  {https://doi.org/10.1103/PhysRevD.39.683} {\bibfield  {journal} {\bibinfo
  {journal} {Phys. Rev. D}\ }\textbf {\bibinfo {volume} {39}},\ \bibinfo
  {pages} {683} (\bibinfo {year} {1989})}\BibitemShut {NoStop}%
\bibitem [{\citenamefont {Kosteleck\'y}\ and\ \citenamefont
  {Potting}(1991)}]{Kostelecky_1991}%
  \BibitemOpen
  \bibfield  {author} {\bibinfo {author} {\bibfnamefont {V.~A.}\ \bibnamefont
  {Kosteleck\'y}}\ and\ \bibinfo {author} {\bibfnamefont {R.}~\bibnamefont
  {Potting}},\ }\bibfield  {title} {\bibinfo {title} {C{PT} and strings},\
  }\href {https://doi.org/https://doi.org/10.1016/0550-3213(91)90071-5}
  {\bibfield  {journal} {\bibinfo  {journal} {Nuclear Physics B}\ }\textbf
  {\bibinfo {volume} {359}},\ \bibinfo {pages} {545} (\bibinfo {year}
  {1991})}\BibitemShut {NoStop}%
\bibitem [{\citenamefont {Kosteleck\'y}\ and\ \citenamefont
  {Russell}(2011)}]{Kostelecky_DataTables}%
  \BibitemOpen
  \bibfield  {author} {\bibinfo {author} {\bibfnamefont {V.~A.}\ \bibnamefont
  {Kosteleck\'y}}\ and\ \bibinfo {author} {\bibfnamefont {N.}~\bibnamefont
  {Russell}},\ }\bibfield  {title} {\bibinfo {title} {Data tables for {L}orentz
  and {CPT} violation},\ }\href {https://doi.org/10.1103/RevModPhys.83.11}
  {\bibfield  {journal} {\bibinfo  {journal} {Rev. Mod. Phys.}\ }\textbf
  {\bibinfo {volume} {83}},\ \bibinfo {pages} {11} (\bibinfo {year}
  {2011})}\BibitemShut {NoStop}%
\bibitem [{\citenamefont {Kosteleck\'y}\ and\ \citenamefont
  {Mewes}(2002{\natexlab{a}})}]{Kostelecky_Mewes_2002}%
  \BibitemOpen
  \bibfield  {author} {\bibinfo {author} {\bibfnamefont {V.~A.}\ \bibnamefont
  {Kosteleck\'y}}\ and\ \bibinfo {author} {\bibfnamefont {M.}~\bibnamefont
  {Mewes}},\ }\bibfield  {title} {\bibinfo {title} {Signals for {L}orentz
  violation in electrodynamics},\ }\href
  {https://doi.org/10.1103/PhysRevD.66.056005} {\bibfield  {journal} {\bibinfo
  {journal} {Phys. Rev. D}\ }\textbf {\bibinfo {volume} {66}},\ \bibinfo
  {pages} {056005} (\bibinfo {year} {2002}{\natexlab{a}})}\BibitemShut
  {NoStop}%
\bibitem [{\citenamefont {Bailey}\ and\ \citenamefont
  {Kosteleck\'y}(2004)}]{Bailey_Kostelecky_2004}%
  \BibitemOpen
  \bibfield  {author} {\bibinfo {author} {\bibfnamefont {Q.~G.}\ \bibnamefont
  {Bailey}}\ and\ \bibinfo {author} {\bibfnamefont {V.~A.}\ \bibnamefont
  {Kosteleck\'y}},\ }\bibfield  {title} {\bibinfo {title} {Lorentz-violating
  electrostatics and magnetostatics},\ }\href
  {https://doi.org/10.1103/PhysRevD.70.076006} {\bibfield  {journal} {\bibinfo
  {journal} {Phys. Rev. D}\ }\textbf {\bibinfo {volume} {70}},\ \bibinfo
  {pages} {076006} (\bibinfo {year} {2004})}\BibitemShut {NoStop}%
\bibitem [{\citenamefont {Mart\'{\i}n-Ruiz}\ and\ \citenamefont
  {Escobar}(2016)}]{Martin_2016}%
  \BibitemOpen
  \bibfield  {author} {\bibinfo {author} {\bibfnamefont {A.}~\bibnamefont
  {Mart\'{\i}n-Ruiz}}\ and\ \bibinfo {author} {\bibfnamefont {C.~A.}\
  \bibnamefont {Escobar}},\ }\bibfield  {title} {\bibinfo {title} {Casimir
  effect between ponderable media as modeled by the standard model extension},\
  }\href {https://doi.org/10.1103/PhysRevD.94.076010} {\bibfield  {journal}
  {\bibinfo  {journal} {Phys. Rev. D}\ }\textbf {\bibinfo {volume} {94}},\
  \bibinfo {pages} {076010} (\bibinfo {year} {2016})}\BibitemShut {NoStop}%
\bibitem [{\citenamefont {Mart\'{\i}n-Ruiz}\ and\ \citenamefont
  {Escobar}(2017)}]{Martin_2017}%
  \BibitemOpen
  \bibfield  {author} {\bibinfo {author} {\bibfnamefont {A.}~\bibnamefont
  {Mart\'{\i}n-Ruiz}}\ and\ \bibinfo {author} {\bibfnamefont {C.~A.}\
  \bibnamefont {Escobar}},\ }\bibfield  {title} {\bibinfo {title} {Local
  effects of the quantum vacuum in {L}orentz-violating electrodynamics},\
  }\href {https://doi.org/10.1103/PhysRevD.95.036011} {\bibfield  {journal}
  {\bibinfo  {journal} {Phys. Rev. D}\ }\textbf {\bibinfo {volume} {95}},\
  \bibinfo {pages} {036011} (\bibinfo {year} {2017})}\BibitemShut {NoStop}%
\bibitem [{\citenamefont {Hohensee}\ \emph {et~al.}(2009)\citenamefont
  {Hohensee}, \citenamefont {Lehnert}, \citenamefont {Phillips},\ and\
  \citenamefont {Walsworth}}]{Hohensee_2009}%
  \BibitemOpen
  \bibfield  {author} {\bibinfo {author} {\bibfnamefont {M.~A.}\ \bibnamefont
  {Hohensee}}, \bibinfo {author} {\bibfnamefont {R.}~\bibnamefont {Lehnert}},
  \bibinfo {author} {\bibfnamefont {D.~F.}\ \bibnamefont {Phillips}},\ and\
  \bibinfo {author} {\bibfnamefont {R.~L.}\ \bibnamefont {Walsworth}},\
  }\bibfield  {title} {\bibinfo {title} {Limits on isotropic {L}orentz
  violation in {QED} from collider physics},\ }\href
  {https://doi.org/10.1103/PhysRevD.80.036010} {\bibfield  {journal} {\bibinfo
  {journal} {Phys. Rev. D}\ }\textbf {\bibinfo {volume} {80}},\ \bibinfo
  {pages} {036010} (\bibinfo {year} {2009})}\BibitemShut {NoStop}%
\bibitem [{\citenamefont {Lehnert}\ and\ \citenamefont
  {Potting}(2004{\natexlab{a}})}]{Lehnert_2004}%
  \BibitemOpen
  \bibfield  {author} {\bibinfo {author} {\bibfnamefont {R.}~\bibnamefont
  {Lehnert}}\ and\ \bibinfo {author} {\bibfnamefont {R.}~\bibnamefont
  {Potting}},\ }\bibfield  {title} {\bibinfo {title} {Vacuum \ifmmode
  \check{C}\else \v{C}\fi{}erenkov radiation},\ }\href
  {https://doi.org/10.1103/PhysRevLett.93.110402} {\bibfield  {journal}
  {\bibinfo  {journal} {Phys. Rev. Lett.}\ }\textbf {\bibinfo {volume} {93}},\
  \bibinfo {pages} {110402} (\bibinfo {year} {2004}{\natexlab{a}})}\BibitemShut
  {NoStop}%
\bibitem [{\citenamefont {Lehnert}\ and\ \citenamefont
  {Potting}(2004{\natexlab{b}})}]{Lehnert_2004_2}%
  \BibitemOpen
  \bibfield  {author} {\bibinfo {author} {\bibfnamefont {R.}~\bibnamefont
  {Lehnert}}\ and\ \bibinfo {author} {\bibfnamefont {R.}~\bibnamefont
  {Potting}},\ }\bibfield  {title} {\bibinfo {title} {\ifmmode \check{C}\else
  \v{C}\fi{}erenkov effect in {L}orentz-violating vacua},\ }\href
  {https://doi.org/10.1103/PhysRevD.70.125010} {\bibfield  {journal} {\bibinfo
  {journal} {Phys. Rev. D}\ }\textbf {\bibinfo {volume} {70}},\ \bibinfo
  {pages} {125010} (\bibinfo {year} {2004}{\natexlab{b}})}\BibitemShut
  {NoStop}%
\bibitem [{\citenamefont {Kosteleck\'y}\ and\ \citenamefont
  {Pickering}(2003)}]{Kostelecky_Pickering_2003}%
  \BibitemOpen
  \bibfield  {author} {\bibinfo {author} {\bibfnamefont {V.~A.}\ \bibnamefont
  {Kosteleck\'y}}\ and\ \bibinfo {author} {\bibfnamefont {A.~G.~M.}\
  \bibnamefont {Pickering}},\ }\bibfield  {title} {\bibinfo {title} {Vacuum
  photon splitting in {L}orentz-violating quantum electrodynamics},\ }\href
  {https://doi.org/10.1103/PhysRevLett.91.031801} {\bibfield  {journal}
  {\bibinfo  {journal} {Phys. Rev. Lett.}\ }\textbf {\bibinfo {volume} {91}},\
  \bibinfo {pages} {031801} (\bibinfo {year} {2003})}\BibitemShut {NoStop}%
\bibitem [{\citenamefont {Colladay}\ \emph {et~al.}(2017)\citenamefont
  {Colladay}, \citenamefont {Noordmans},\ and\ \citenamefont
  {Potting}}]{Colladay_2017}%
  \BibitemOpen
  \bibfield  {author} {\bibinfo {author} {\bibfnamefont {D.}~\bibnamefont
  {Colladay}}, \bibinfo {author} {\bibfnamefont {J.~P.}\ \bibnamefont
  {Noordmans}},\ and\ \bibinfo {author} {\bibfnamefont {R.}~\bibnamefont
  {Potting}},\ }\bibfield  {title} {\bibinfo {title} {Cosmic-ray fermion decay
  by emission of on-shell {W} bosons with {CPT} violation},\ }\href
  {https://doi.org/10.1103/PhysRevD.96.035034} {\bibfield  {journal} {\bibinfo
  {journal} {Phys. Rev. D}\ }\textbf {\bibinfo {volume} {96}},\ \bibinfo
  {pages} {035034} (\bibinfo {year} {2017})}\BibitemShut {NoStop}%
\bibitem [{\citenamefont {Mart\'{\i}n-Ruiz}\ \emph
  {et~al.}(2015{\natexlab{a}})\citenamefont {Mart\'{\i}n-Ruiz}, \citenamefont
  {Cambiaso},\ and\ \citenamefont {Urrutia}}]{Martin_TI_2015}%
  \BibitemOpen
  \bibfield  {author} {\bibinfo {author} {\bibfnamefont {A.}~\bibnamefont
  {Mart\'{\i}n-Ruiz}}, \bibinfo {author} {\bibfnamefont {M.}~\bibnamefont
  {Cambiaso}},\ and\ \bibinfo {author} {\bibfnamefont {L.~F.}\ \bibnamefont
  {Urrutia}},\ }\bibfield  {title} {\bibinfo {title} {Green's function approach
  to {C}hern-{S}imons extended electrodynamics: An effective theory describing
  topological insulators},\ }\href {https://doi.org/10.1103/PhysRevD.92.125015}
  {\bibfield  {journal} {\bibinfo  {journal} {Phys. Rev. D}\ }\textbf {\bibinfo
  {volume} {92}},\ \bibinfo {pages} {125015} (\bibinfo {year}
  {2015}{\natexlab{a}})}\BibitemShut {NoStop}%
\bibitem [{\citenamefont {Mart\'{\i}n-Ruiz}\ \emph
  {et~al.}(2016{\natexlab{a}})\citenamefont {Mart\'{\i}n-Ruiz}, \citenamefont
  {Cambiaso},\ and\ \citenamefont {Urrutia}}]{Martin_TI_2016}%
  \BibitemOpen
  \bibfield  {author} {\bibinfo {author} {\bibfnamefont {A.}~\bibnamefont
  {Mart\'{\i}n-Ruiz}}, \bibinfo {author} {\bibfnamefont {M.}~\bibnamefont
  {Cambiaso}},\ and\ \bibinfo {author} {\bibfnamefont {L.~F.}\ \bibnamefont
  {Urrutia}},\ }\bibfield  {title} {\bibinfo {title} {Electro- and
  magnetostatics of topological insulators as modeled by planar, spherical, and
  cylindrical $\ensuremath{\theta}$ boundaries: {G}reen's function approach},\
  }\href {https://doi.org/10.1103/PhysRevD.93.045022} {\bibfield  {journal}
  {\bibinfo  {journal} {Phys. Rev. D}\ }\textbf {\bibinfo {volume} {93}},\
  \bibinfo {pages} {045022} (\bibinfo {year} {2016}{\natexlab{a}})}\BibitemShut
  {NoStop}%
\bibitem [{\citenamefont {Mart\'{\i}n-Ruiz}\ \emph
  {et~al.}(2016{\natexlab{b}})\citenamefont {Mart\'{\i}n-Ruiz}, \citenamefont
  {Cambiaso},\ and\ \citenamefont {Urrutia}}]{Martin_TI_2016_2}%
  \BibitemOpen
  \bibfield  {author} {\bibinfo {author} {\bibfnamefont {A.}~\bibnamefont
  {Mart\'{\i}n-Ruiz}}, \bibinfo {author} {\bibfnamefont {M.}~\bibnamefont
  {Cambiaso}},\ and\ \bibinfo {author} {\bibfnamefont {L.~F.}\ \bibnamefont
  {Urrutia}},\ }\bibfield  {title} {\bibinfo {title} {Electromagnetic
  description of three-dimensional time-reversal invariant ponderable
  topological insulators},\ }\href {https://doi.org/10.1103/PhysRevD.94.085019}
  {\bibfield  {journal} {\bibinfo  {journal} {Phys. Rev. D}\ }\textbf {\bibinfo
  {volume} {94}},\ \bibinfo {pages} {085019} (\bibinfo {year}
  {2016}{\natexlab{b}})}\BibitemShut {NoStop}%
\bibitem [{\citenamefont {Mart\'{\i}n-Ruiz}\ \emph {et~al.}(2019)\citenamefont
  {Mart\'{\i}n-Ruiz}, \citenamefont {Cambiaso},\ and\ \citenamefont
  {Urrutia}}]{Martin_WS_2019}%
  \BibitemOpen
  \bibfield  {author} {\bibinfo {author} {\bibfnamefont {A.}~\bibnamefont
  {Mart\'{\i}n-Ruiz}}, \bibinfo {author} {\bibfnamefont {M.}~\bibnamefont
  {Cambiaso}},\ and\ \bibinfo {author} {\bibfnamefont {L.~F.}\ \bibnamefont
  {Urrutia}},\ }\bibfield  {title} {\bibinfo {title} {Electromagnetic fields
  induced by an electric charge near a {W}eyl semimetal},\ }\href
  {https://doi.org/10.1103/PhysRevB.99.155142} {\bibfield  {journal} {\bibinfo
  {journal} {Phys. Rev. B}\ }\textbf {\bibinfo {volume} {99}},\ \bibinfo
  {pages} {155142} (\bibinfo {year} {2019})}\BibitemShut {NoStop}%
\bibitem [{\citenamefont {Grushin}(2012)}]{Grushin_2012}%
  \BibitemOpen
  \bibfield  {author} {\bibinfo {author} {\bibfnamefont {A.~G.}\ \bibnamefont
  {Grushin}},\ }\bibfield  {title} {\bibinfo {title} {Consequences of a
  condensed matter realization of {L}orentz-violating {QED} in {W}eyl
  semi-metals},\ }\href {https://doi.org/10.1103/PhysRevD.86.045001} {\bibfield
   {journal} {\bibinfo  {journal} {Phys. Rev. D}\ }\textbf {\bibinfo {volume}
  {86}},\ \bibinfo {pages} {045001} (\bibinfo {year} {2012})}\BibitemShut
  {NoStop}%
\bibitem [{\citenamefont {Gómez}\ \emph {et~al.}(2022)\citenamefont {Gómez},
  \citenamefont {Martín-Ruiz},\ and\ \citenamefont {Urrutia}}]{Gomez_2022}%
  \BibitemOpen
  \bibfield  {author} {\bibinfo {author} {\bibfnamefont {A.}~\bibnamefont
  {Gómez}}, \bibinfo {author} {\bibfnamefont {A.}~\bibnamefont
  {Martín-Ruiz}},\ and\ \bibinfo {author} {\bibfnamefont {L.~F.}\ \bibnamefont
  {Urrutia}},\ }\bibfield  {title} {\bibinfo {title} {Effective electromagnetic
  actions for {L}orentz violating theories exhibiting the axial anomaly},\
  }\href {https://doi.org/https://doi.org/10.1016/j.physletb.2022.137043}
  {\bibfield  {journal} {\bibinfo  {journal} {Physics Letters B}\ }\textbf
  {\bibinfo {volume} {829}},\ \bibinfo {pages} {137043} (\bibinfo {year}
  {2022})}\BibitemShut {NoStop}%
\bibitem [{\citenamefont {Kosteleck\'y}\ \emph {et~al.}(2022)\citenamefont
  {Kosteleck\'y}, \citenamefont {Lehnert}, \citenamefont {McGinnis},
  \citenamefont {Schreck},\ and\ \citenamefont {Seradjeh}}]{Kostelecky_2022}%
  \BibitemOpen
  \bibfield  {author} {\bibinfo {author} {\bibfnamefont {V.~A.}\ \bibnamefont
  {Kosteleck\'y}}, \bibinfo {author} {\bibfnamefont {R.}~\bibnamefont
  {Lehnert}}, \bibinfo {author} {\bibfnamefont {N.}~\bibnamefont {McGinnis}},
  \bibinfo {author} {\bibfnamefont {M.}~\bibnamefont {Schreck}},\ and\ \bibinfo
  {author} {\bibfnamefont {B.}~\bibnamefont {Seradjeh}},\ }\bibfield  {title}
  {\bibinfo {title} {Lorentz violation in {D}irac and {W}eyl semimetals},\
  }\href {https://doi.org/10.1103/PhysRevResearch.4.023106} {\bibfield
  {journal} {\bibinfo  {journal} {Phys. Rev. Research}\ }\textbf {\bibinfo
  {volume} {4}},\ \bibinfo {pages} {023106} (\bibinfo {year}
  {2022})}\BibitemShut {NoStop}%
\bibitem [{\citenamefont {Edwards}\ and\ \citenamefont
  {Kostelecký}(2018)}]{Kostelecky-Edwards_2018}%
  \BibitemOpen
  \bibfield  {author} {\bibinfo {author} {\bibfnamefont {B.~R.}\ \bibnamefont
  {Edwards}}\ and\ \bibinfo {author} {\bibfnamefont {V.~A.}\ \bibnamefont
  {Kostelecký}},\ }\bibfield  {title} {\bibinfo {title} {Riemann-{F}insler
  geometry and {L}orentz-violating scalar fields},\ }\href
  {https://doi.org/https://doi.org/10.1016/j.physletb.2018.10.011} {\bibfield
  {journal} {\bibinfo  {journal} {Physics Letters B}\ }\textbf {\bibinfo
  {volume} {786}},\ \bibinfo {pages} {319} (\bibinfo {year}
  {2018})}\BibitemShut {NoStop}%
\bibitem [{\citenamefont {Escobar}\ \emph
  {et~al.}(2020{\natexlab{a}})\citenamefont {Escobar}, \citenamefont {Medel},\
  and\ \citenamefont {Mart\'{\i}n-Ruiz}}]{Medel_2020}%
  \BibitemOpen
  \bibfield  {author} {\bibinfo {author} {\bibfnamefont {C.~A.}\ \bibnamefont
  {Escobar}}, \bibinfo {author} {\bibfnamefont {L.}~\bibnamefont {Medel}},\
  and\ \bibinfo {author} {\bibfnamefont {A.}~\bibnamefont {Mart\'{\i}n-Ruiz}},\
  }\bibfield  {title} {\bibinfo {title} {Casimir effect in {L}orentz-violating
  scalar field theory: A local approach},\ }\href
  {https://doi.org/10.1103/PhysRevD.101.095011} {\bibfield  {journal} {\bibinfo
   {journal} {Phys. Rev. D}\ }\textbf {\bibinfo {volume} {101}},\ \bibinfo
  {pages} {095011} (\bibinfo {year} {2020}{\natexlab{a}})}\BibitemShut
  {NoStop}%
\bibitem [{\citenamefont {Mart\'{\i}n-Ruiz}\ \emph {et~al.}(2020)\citenamefont
  {Mart\'{\i}n-Ruiz}, \citenamefont {Escobar}, \citenamefont {Escobar-Ruiz},\
  and\ \citenamefont {Franca}}]{Martin-Ruiz_2020}%
  \BibitemOpen
  \bibfield  {author} {\bibinfo {author} {\bibfnamefont {A.}~\bibnamefont
  {Mart\'{\i}n-Ruiz}}, \bibinfo {author} {\bibfnamefont {C.~A.}\ \bibnamefont
  {Escobar}}, \bibinfo {author} {\bibfnamefont {A.~M.}\ \bibnamefont
  {Escobar-Ruiz}},\ and\ \bibinfo {author} {\bibfnamefont {O.~J.}\ \bibnamefont
  {Franca}},\ }\bibfield  {title} {\bibinfo {title} {Lorentz violating scalar
  {C}asimir effect for a ${D}$-dimensional sphere},\ }\href
  {https://doi.org/10.1103/PhysRevD.102.015027} {\bibfield  {journal} {\bibinfo
   {journal} {Phys. Rev. D}\ }\textbf {\bibinfo {volume} {102}},\ \bibinfo
  {pages} {015027} (\bibinfo {year} {2020})}\BibitemShut {NoStop}%
\bibitem [{\citenamefont {Escobar}\ \emph
  {et~al.}(2020{\natexlab{b}})\citenamefont {Escobar}, \citenamefont
  {Martín-Ruiz}, \citenamefont {Franca},\ and\ \citenamefont
  {Garcia}}]{Escobar_2020}%
  \BibitemOpen
  \bibfield  {author} {\bibinfo {author} {\bibfnamefont {C.~A.}\ \bibnamefont
  {Escobar}}, \bibinfo {author} {\bibfnamefont {A.}~\bibnamefont
  {Martín-Ruiz}}, \bibinfo {author} {\bibfnamefont {O.~J.}\ \bibnamefont
  {Franca}},\ and\ \bibinfo {author} {\bibfnamefont {M.~A.~G.}\ \bibnamefont
  {Garcia}},\ }\bibfield  {title} {\bibinfo {title} {A non-perturbative
  approach to the scalar {C}asimir effect with {L}orentz symmetry violation},\
  }\href {https://doi.org/https://doi.org/10.1016/j.physletb.2020.135567}
  {\bibfield  {journal} {\bibinfo  {journal} {Physics Letters B}\ }\textbf
  {\bibinfo {volume} {807}},\ \bibinfo {pages} {135567} (\bibinfo {year}
  {2020}{\natexlab{b}})}\BibitemShut {NoStop}%
\bibitem [{\citenamefont {Escobar-Ruiz}\ \emph {et~al.}(2021)\citenamefont
  {Escobar-Ruiz}, \citenamefont {Mart\'{\i}n-Ruiz}, \citenamefont {Escobar},\
  and\ \citenamefont {Linares}}]{Escobar-Ruiz_2021}%
  \BibitemOpen
  \bibfield  {author} {\bibinfo {author} {\bibfnamefont {A.~M.}\ \bibnamefont
  {Escobar-Ruiz}}, \bibinfo {author} {\bibfnamefont {A.}~\bibnamefont
  {Mart\'{\i}n-Ruiz}}, \bibinfo {author} {\bibfnamefont {C.~A.}\ \bibnamefont
  {Escobar}},\ and\ \bibinfo {author} {\bibfnamefont {R.}~\bibnamefont
  {Linares}},\ }\bibfield  {title} {\bibinfo {title} {Scalar {C}asimir effect
  for a conducting cylinder in a {L}orentz-violating background},\ }\href
  {https://doi.org/10.1142/S0217751X21501682} {\bibfield  {journal} {\bibinfo
  {journal} {International Journal of Modern Physics A}\ }\textbf {\bibinfo
  {volume} {36}},\ \bibinfo {pages} {2150168} (\bibinfo {year}
  {2021})}\BibitemShut {NoStop}%
\bibitem [{\citenamefont {Furtado}\ \emph {et~al.}(2021)\citenamefont
  {Furtado}, \citenamefont {Costa~Filho}, \citenamefont {Morais},\ and\
  \citenamefont {Jardim}}]{Furtado_2021}%
  \BibitemOpen
  \bibfield  {author} {\bibinfo {author} {\bibfnamefont {J.}~\bibnamefont
  {Furtado}}, \bibinfo {author} {\bibfnamefont {R.~M.~M.}\ \bibnamefont
  {Costa~Filho}}, \bibinfo {author} {\bibfnamefont {A.~F.}\ \bibnamefont
  {Morais}},\ and\ \bibinfo {author} {\bibfnamefont {I.~C.}\ \bibnamefont
  {Jardim}},\ }\bibfield  {title} {\bibinfo {title} {Effects of {L}orentz
  violation in superconductivity},\ }\href
  {https://doi.org/10.1209/0295-5075/ac36f0} {\bibfield  {journal} {\bibinfo
  {journal} {Europhysics Letters}\ }\textbf {\bibinfo {volume} {136}},\
  \bibinfo {pages} {51001} (\bibinfo {year} {2021})}\BibitemShut {NoStop}%
\bibitem [{\citenamefont {Tian}\ and\ \citenamefont {Du}(2021)}]{Tian_2021}%
  \BibitemOpen
  \bibfield  {author} {\bibinfo {author} {\bibfnamefont {Z.}~\bibnamefont
  {Tian}}\ and\ \bibinfo {author} {\bibfnamefont {J.}~\bibnamefont {Du}},\
  }\bibfield  {title} {\bibinfo {title} {Probing low-energy {L}orentz violation
  from high-energy modified dispersion in dipolar {B}ose-{E}instein
  condensates},\ }\href {https://doi.org/10.1103/PhysRevD.103.085014}
  {\bibfield  {journal} {\bibinfo  {journal} {Phys. Rev. D}\ }\textbf {\bibinfo
  {volume} {103}},\ \bibinfo {pages} {085014} (\bibinfo {year}
  {2021})}\BibitemShut {NoStop}%
\bibitem [{\citenamefont {Aguirre}\ \emph {et~al.}(2021)\citenamefont
  {Aguirre}, \citenamefont {Flores-Hidalgo}, \citenamefont {Rana},\ and\
  \citenamefont {Souza}}]{Aguirre_2021}%
  \BibitemOpen
  \bibfield  {author} {\bibinfo {author} {\bibfnamefont {A.~R.}\ \bibnamefont
  {Aguirre}}, \bibinfo {author} {\bibfnamefont {G.}~\bibnamefont
  {Flores-Hidalgo}}, \bibinfo {author} {\bibfnamefont {R.~G.}\ \bibnamefont
  {Rana}},\ and\ \bibinfo {author} {\bibfnamefont {E.~S.}\ \bibnamefont
  {Souza}},\ }\bibfield  {title} {\bibinfo {title} {The {L}orentz-violating
  real scalar field at thermal equilibrium},\ }\href
  {https://doi.org/10.1140/epjc/s10052-021-09250-1} {\bibfield  {journal}
  {\bibinfo  {journal} {The European Physical Journal C}\ }\textbf {\bibinfo
  {volume} {81}},\ \bibinfo {pages} {459} (\bibinfo {year} {2021})}\BibitemShut
  {NoStop}%
\bibitem [{\citenamefont {Filho}\ and\ \citenamefont
  {Reis}(2021)}]{Filho_2021}%
  \BibitemOpen
  \bibfield  {author} {\bibinfo {author} {\bibfnamefont {A.~A.~A.}\
  \bibnamefont {Filho}}\ and\ \bibinfo {author} {\bibfnamefont {J.~A. A.~S.}\
  \bibnamefont {Reis}},\ }\bibfield  {title} {\bibinfo {title} {Thermal aspects
  of interacting quantum gases in {L}orentz-violating scenarios},\ }\href
  {https://doi.org/10.1140/epjp/s13360-021-01289-z} {\bibfield  {journal}
  {\bibinfo  {journal} {The European Physical Journal Plus}\ }\textbf {\bibinfo
  {volume} {136}},\ \bibinfo {pages} {310} (\bibinfo {year}
  {2021})}\BibitemShut {NoStop}%
\bibitem [{\citenamefont {Nesvizhevsky}\ \emph {et~al.}(2002)\citenamefont
  {Nesvizhevsky}, \citenamefont {Börner}, \citenamefont {Petukhov},
  \citenamefont {Abele}, \citenamefont {Baeßler}, \citenamefont {Rueß},
  \citenamefont {Stöferle}, \citenamefont {Westphal}, \citenamefont
  {Gagarski}, \citenamefont {Petrov},\ and\ \citenamefont
  {Strelkov}}]{Nesvizhevsky_2002}%
  \BibitemOpen
  \bibfield  {author} {\bibinfo {author} {\bibfnamefont {V.~V.}\ \bibnamefont
  {Nesvizhevsky}}, \bibinfo {author} {\bibfnamefont {H.~G.}\ \bibnamefont
  {Börner}}, \bibinfo {author} {\bibfnamefont {A.~K.}\ \bibnamefont
  {Petukhov}}, \bibinfo {author} {\bibfnamefont {H.}~\bibnamefont {Abele}},
  \bibinfo {author} {\bibfnamefont {S.}~\bibnamefont {Baeßler}}, \bibinfo
  {author} {\bibfnamefont {F.~J.}\ \bibnamefont {Rueß}}, \bibinfo {author}
  {\bibfnamefont {T.}~\bibnamefont {Stöferle}}, \bibinfo {author}
  {\bibfnamefont {A.}~\bibnamefont {Westphal}}, \bibinfo {author}
  {\bibfnamefont {A.~M.}\ \bibnamefont {Gagarski}}, \bibinfo {author}
  {\bibfnamefont {G.~A.}\ \bibnamefont {Petrov}},\ and\ \bibinfo {author}
  {\bibfnamefont {A.~V.}\ \bibnamefont {Strelkov}},\ }\bibfield  {title}
  {\bibinfo {title} {Quantum states of neutrons in the {E}arth's gravitational
  field},\ }\href {https://doi.org/10.1038/415297a} {\bibfield  {journal}
  {\bibinfo  {journal} {Nature}\ }\textbf {\bibinfo {volume} {415}},\ \bibinfo
  {pages} {297} (\bibinfo {year} {2002})}\BibitemShut {NoStop}%
\bibitem [{\citenamefont {Cronenberg}\ \emph {et~al.}(2018)\citenamefont
  {Cronenberg}, \citenamefont {Brax}, \citenamefont {Filter}, \citenamefont
  {Geltenbort}, \citenamefont {Jenke}, \citenamefont {Pignol}, \citenamefont
  {Pitschmann}, \citenamefont {Thalhammer},\ and\ \citenamefont
  {Abele}}]{Cronenberg_2018}%
  \BibitemOpen
  \bibfield  {author} {\bibinfo {author} {\bibfnamefont {G.}~\bibnamefont
  {Cronenberg}}, \bibinfo {author} {\bibfnamefont {P.}~\bibnamefont {Brax}},
  \bibinfo {author} {\bibfnamefont {H.}~\bibnamefont {Filter}}, \bibinfo
  {author} {\bibfnamefont {P.}~\bibnamefont {Geltenbort}}, \bibinfo {author}
  {\bibfnamefont {T.}~\bibnamefont {Jenke}}, \bibinfo {author} {\bibfnamefont
  {G.}~\bibnamefont {Pignol}}, \bibinfo {author} {\bibfnamefont
  {M.}~\bibnamefont {Pitschmann}}, \bibinfo {author} {\bibfnamefont
  {M.}~\bibnamefont {Thalhammer}},\ and\ \bibinfo {author} {\bibfnamefont
  {H.}~\bibnamefont {Abele}},\ }\bibfield  {title} {\bibinfo {title} {Acoustic
  {R}abi oscillations between gravitational quantum states and impact on
  symmetron dark energy},\ }\href {https://doi.org/10.1038/s41567-018-0205-x}
  {\bibfield  {journal} {\bibinfo  {journal} {Nature Physics}\ }\textbf
  {\bibinfo {volume} {14}},\ \bibinfo {pages} {1022} (\bibinfo {year}
  {2018})}\BibitemShut {NoStop}%
\bibitem [{\citenamefont {Colella}\ \emph {et~al.}(1975)\citenamefont
  {Colella}, \citenamefont {Overhauser},\ and\ \citenamefont
  {Werner}}]{COW_1975}%
  \BibitemOpen
  \bibfield  {author} {\bibinfo {author} {\bibfnamefont {R.}~\bibnamefont
  {Colella}}, \bibinfo {author} {\bibfnamefont {A.~W.}\ \bibnamefont
  {Overhauser}},\ and\ \bibinfo {author} {\bibfnamefont {S.~A.}\ \bibnamefont
  {Werner}},\ }\bibfield  {title} {\bibinfo {title} {Observation of
  gravitationally induced quantum interference},\ }\href
  {https://doi.org/10.1103/PhysRevLett.34.1472} {\bibfield  {journal} {\bibinfo
   {journal} {Phys. Rev. Lett.}\ }\textbf {\bibinfo {volume} {34}},\ \bibinfo
  {pages} {1472} (\bibinfo {year} {1975})}\BibitemShut {NoStop}%
\bibitem [{\citenamefont {Kosteleck\'y}\ and\ \citenamefont
  {Lehnert}(2001)}]{Kosteleck_Lehnert_2001}%
  \BibitemOpen
  \bibfield  {author} {\bibinfo {author} {\bibfnamefont {V.~A.}\ \bibnamefont
  {Kosteleck\'y}}\ and\ \bibinfo {author} {\bibfnamefont {R.}~\bibnamefont
  {Lehnert}},\ }\bibfield  {title} {\bibinfo {title} {Stability, causality, and
  {L}orentz and {CPT} violation},\ }\href
  {https://doi.org/10.1103/PhysRevD.63.065008} {\bibfield  {journal} {\bibinfo
  {journal} {Phys. Rev. D}\ }\textbf {\bibinfo {volume} {63}},\ \bibinfo
  {pages} {065008} (\bibinfo {year} {2001})}\BibitemShut {NoStop}%
\bibitem [{\citenamefont {Chang}\ and\ \citenamefont
  {Wang}(2012)}]{Chang_2012}%
  \BibitemOpen
  \bibfield  {author} {\bibinfo {author} {\bibfnamefont {Z.}~\bibnamefont
  {Chang}}\ and\ \bibinfo {author} {\bibfnamefont {S.}~\bibnamefont {Wang}},\
  }\bibfield  {title} {\bibinfo {title} {Lorentz invariance violation and
  electromagnetic field in an intrinsically anisotropic spacetime},\ }\href
  {https://doi.org/10.1140/epjc/s10052-012-2165-0} {\bibfield  {journal}
  {\bibinfo  {journal} {The European Physical Journal C}\ }\textbf {\bibinfo
  {volume} {72}},\ \bibinfo {pages} {2165} (\bibinfo {year}
  {2012})}\BibitemShut {NoStop}%
\bibitem [{\citenamefont {Zhang}(2017)}]{Zhang_2017}%
  \BibitemOpen
  \bibfield  {author} {\bibinfo {author} {\bibfnamefont {A.}~\bibnamefont
  {Zhang}},\ }\bibfield  {title} {\bibinfo {title} {Theoretical analysis of
  {C}asimir and thermal {C}asimir effect in stationary space-time},\ }\href
  {https://doi.org/https://doi.org/10.1016/j.physletb.2017.08.012} {\bibfield
  {journal} {\bibinfo  {journal} {Physics Letters B}\ }\textbf {\bibinfo
  {volume} {773}},\ \bibinfo {pages} {125} (\bibinfo {year}
  {2017})}\BibitemShut {NoStop}%
\bibitem [{\citenamefont {Mart\'{\i}n-Ruiz}\ and\ \citenamefont
  {Escobar}(2018)}]{Martin_2018}%
  \BibitemOpen
  \bibfield  {author} {\bibinfo {author} {\bibfnamefont {A.}~\bibnamefont
  {Mart\'{\i}n-Ruiz}}\ and\ \bibinfo {author} {\bibfnamefont {C.~A.}\
  \bibnamefont {Escobar}},\ }\bibfield  {title} {\bibinfo {title} {Testing
  {L}orentz and {CPT} invariance with ultracold neutrons},\ }\href
  {https://doi.org/10.1103/PhysRevD.97.095039} {\bibfield  {journal} {\bibinfo
  {journal} {Phys. Rev. D}\ }\textbf {\bibinfo {volume} {97}},\ \bibinfo
  {pages} {095039} (\bibinfo {year} {2018})}\BibitemShut {NoStop}%
\bibitem [{\citenamefont {Xiao}\ and\ \citenamefont {Shao}(2020)}]{Xiao_2020}%
  \BibitemOpen
  \bibfield  {author} {\bibinfo {author} {\bibfnamefont {Z.}~\bibnamefont
  {Xiao}}\ and\ \bibinfo {author} {\bibfnamefont {L.}~\bibnamefont {Shao}},\
  }\bibfield  {title} {\bibinfo {title} {The {CPT}-violating effects on
  neutrons' gravitational bound state},\ }\href
  {https://doi.org/10.1088/1361-6471/ab8c30} {\bibfield  {journal} {\bibinfo
  {journal} {Journal of Physics G: Nuclear and Particle Physics}\ }\textbf
  {\bibinfo {volume} {47}},\ \bibinfo {pages} {085002} (\bibinfo {year}
  {2020})}\BibitemShut {NoStop}%
\bibitem [{\citenamefont {Ivanov}\ \emph {et~al.}(2019)\citenamefont {Ivanov},
  \citenamefont {Wellenzohn},\ and\ \citenamefont {Abele}}]{Ivanov_2019}%
  \BibitemOpen
  \bibfield  {author} {\bibinfo {author} {\bibfnamefont {A.~N.}\ \bibnamefont
  {Ivanov}}, \bibinfo {author} {\bibfnamefont {M.}~\bibnamefont {Wellenzohn}},\
  and\ \bibinfo {author} {\bibfnamefont {H.}~\bibnamefont {Abele}},\ }\bibfield
   {title} {\bibinfo {title} {Probing of violation of {L}orentz invariance by
  ultracold neutrons in the standard model extension},\ }\href
  {https://doi.org/https://doi.org/10.1016/j.physletb.2019.134819} {\bibfield
  {journal} {\bibinfo  {journal} {Physics Letters B}\ }\textbf {\bibinfo
  {volume} {797}},\ \bibinfo {pages} {134819} (\bibinfo {year}
  {2019})}\BibitemShut {NoStop}%
\bibitem [{\citenamefont {Escobar}\ and\ \citenamefont
  {Mart\'{\i}n-Ruiz}(2019)}]{Escobar_2019}%
  \BibitemOpen
  \bibfield  {author} {\bibinfo {author} {\bibfnamefont {C.~A.}\ \bibnamefont
  {Escobar}}\ and\ \bibinfo {author} {\bibfnamefont {A.}~\bibnamefont
  {Mart\'{\i}n-Ruiz}},\ }\bibfield  {title} {\bibinfo {title} {Gravitational
  searches for {L}orentz violation with ultracold neutrons},\ }\href
  {https://doi.org/10.1103/PhysRevD.99.075032} {\bibfield  {journal} {\bibinfo
  {journal} {Phys. Rev. D}\ }\textbf {\bibinfo {volume} {99}},\ \bibinfo
  {pages} {075032} (\bibinfo {year} {2019})}\BibitemShut {NoStop}%
\bibitem [{\citenamefont {Ivanov}\ \emph {et~al.}(2021)\citenamefont {Ivanov},
  \citenamefont {Wellenzohn},\ and\ \citenamefont {Abele}}]{Ivanov_2021}%
  \BibitemOpen
  \bibfield  {author} {\bibinfo {author} {\bibfnamefont {A.~N.}\ \bibnamefont
  {Ivanov}}, \bibinfo {author} {\bibfnamefont {M.}~\bibnamefont {Wellenzohn}},\
  and\ \bibinfo {author} {\bibfnamefont {H.}~\bibnamefont {Abele}},\ }\bibfield
   {title} {\bibinfo {title} {Quantum gravitational states of ultracold
  neutrons as a tool for probing of beyond-{R}iemann gravity},\ }\href
  {https://doi.org/https://doi.org/10.1016/j.physletb.2021.136640} {\bibfield
  {journal} {\bibinfo  {journal} {Physics Letters B}\ }\textbf {\bibinfo
  {volume} {822}},\ \bibinfo {pages} {136640} (\bibinfo {year}
  {2021})}\BibitemShut {NoStop}%
\bibitem [{\citenamefont {Vallée}\ and\ \citenamefont
  {Soares}(2004)}]{Vallee_2004}%
  \BibitemOpen
  \bibfield  {author} {\bibinfo {author} {\bibfnamefont {O.}~\bibnamefont
  {Vallée}}\ and\ \bibinfo {author} {\bibfnamefont {M.}~\bibnamefont
  {Soares}},\ }\href {https://doi.org/10.1142/p345} {\emph {\bibinfo {title}
  {Airy Functions and Applications to Physics}}}\ (\bibinfo  {publisher}
  {Imperial College Press},\ \bibinfo {year} {2004})\BibitemShut {NoStop}%
\bibitem [{\citenamefont {Nesvizhevsky}\ \emph {et~al.}(2005)\citenamefont
  {Nesvizhevsky}, \citenamefont {Petukhov}, \citenamefont {Börner},
  \citenamefont {Baranova}, \citenamefont {Gagarski}, \citenamefont {Petrov},
  \citenamefont {Protasov}, \citenamefont {Voronin}, \citenamefont {Baeßler},
  \citenamefont {Abele}, \citenamefont {Westphal},\ and\ \citenamefont
  {Lucovac}}]{Nesvizhevsky_2005}%
  \BibitemOpen
  \bibfield  {author} {\bibinfo {author} {\bibfnamefont {V.~V.}\ \bibnamefont
  {Nesvizhevsky}}, \bibinfo {author} {\bibfnamefont {A.~K.}\ \bibnamefont
  {Petukhov}}, \bibinfo {author} {\bibfnamefont {H.~G.}\ \bibnamefont
  {Börner}}, \bibinfo {author} {\bibfnamefont {T.~A.}\ \bibnamefont
  {Baranova}}, \bibinfo {author} {\bibfnamefont {A.~M.}\ \bibnamefont
  {Gagarski}}, \bibinfo {author} {\bibfnamefont {G.~A.}\ \bibnamefont
  {Petrov}}, \bibinfo {author} {\bibfnamefont {K.~V.}\ \bibnamefont
  {Protasov}}, \bibinfo {author} {\bibfnamefont {A.~Y.}\ \bibnamefont
  {Voronin}}, \bibinfo {author} {\bibfnamefont {S.}~\bibnamefont {Baeßler}},
  \bibinfo {author} {\bibfnamefont {H.}~\bibnamefont {Abele}}, \bibinfo
  {author} {\bibfnamefont {A.}~\bibnamefont {Westphal}},\ and\ \bibinfo
  {author} {\bibfnamefont {L.}~\bibnamefont {Lucovac}},\ }\bibfield  {title}
  {\bibinfo {title} {Study of the neutron quantum states in the gravity
  field},\ }\href {https://doi.org/10.1140/epjc/s2005-02135-y} {\bibfield
  {journal} {\bibinfo  {journal} {The European Physical Journal C}\ }\textbf
  {\bibinfo {volume} {40}},\ \bibinfo {pages} {479} (\bibinfo {year}
  {2005})}\BibitemShut {NoStop}%
\bibitem [{\citenamefont {Nesvizhevsky}\ \emph {et~al.}(2008)\citenamefont
  {Nesvizhevsky}, \citenamefont {Pignol},\ and\ \citenamefont
  {Protasov}}]{Nesvizhevsky_2008}%
  \BibitemOpen
  \bibfield  {author} {\bibinfo {author} {\bibfnamefont {V.~V.}\ \bibnamefont
  {Nesvizhevsky}}, \bibinfo {author} {\bibfnamefont {G.}~\bibnamefont
  {Pignol}},\ and\ \bibinfo {author} {\bibfnamefont {K.~V.}\ \bibnamefont
  {Protasov}},\ }\bibfield  {title} {\bibinfo {title} {Neutron scattering and
  extra-short-range interactions},\ }\href
  {https://doi.org/10.1103/PhysRevD.77.034020} {\bibfield  {journal} {\bibinfo
  {journal} {Phys. Rev. D}\ }\textbf {\bibinfo {volume} {77}},\ \bibinfo
  {pages} {034020} (\bibinfo {year} {2008})}\BibitemShut {NoStop}%
\bibitem [{\citenamefont {Baeßler}\ \emph {et~al.}(2009)\citenamefont
  {Baeßler}, \citenamefont {NesvizhevskY}, \citenamefont {Pignol},
  \citenamefont {Protasov},\ and\ \citenamefont {Voronin}}]{Baeler_2009}%
  \BibitemOpen
  \bibfield  {author} {\bibinfo {author} {\bibfnamefont {S.}~\bibnamefont
  {Baeßler}}, \bibinfo {author} {\bibfnamefont {V.~V.}\ \bibnamefont
  {NesvizhevskY}}, \bibinfo {author} {\bibfnamefont {G.}~\bibnamefont
  {Pignol}}, \bibinfo {author} {\bibfnamefont {K.~V.}\ \bibnamefont
  {Protasov}},\ and\ \bibinfo {author} {\bibfnamefont {A.~Y.}\ \bibnamefont
  {Voronin}},\ }\bibfield  {title} {\bibinfo {title} {Constraints on
  spin-dependent short-range interactions using gravitational quantum levels of
  ultracold neutrons},\ }\href
  {https://doi.org/https://doi.org/10.1016/j.nima.2009.07.048} {\bibfield
  {journal} {\bibinfo  {journal} {Nuclear Instruments and Methods in Physics
  Research Section A: Accelerators, Spectrometers, Detectors and Associated
  Equipment}\ }\textbf {\bibinfo {volume} {611}},\ \bibinfo {pages} {149}
  (\bibinfo {year} {2009})}\BibitemShut {NoStop}%
\bibitem [{\citenamefont {Antoniadis}\ \emph {et~al.}(2011)\citenamefont
  {Antoniadis}, \citenamefont {Baessler}, \citenamefont {Büchner},
  \citenamefont {Fedorov}, \citenamefont {Hoedl}, \citenamefont {Lambrecht},
  \citenamefont {Nesvizhevsky}, \citenamefont {Pignol}, \citenamefont
  {Protasov}, \citenamefont {Reynaud},\ and\ \citenamefont
  {Sobolev}}]{Antoniadis_2011}%
  \BibitemOpen
  \bibfield  {author} {\bibinfo {author} {\bibfnamefont {I.}~\bibnamefont
  {Antoniadis}}, \bibinfo {author} {\bibfnamefont {S.}~\bibnamefont
  {Baessler}}, \bibinfo {author} {\bibfnamefont {M.}~\bibnamefont {Büchner}},
  \bibinfo {author} {\bibfnamefont {V.~V.}\ \bibnamefont {Fedorov}}, \bibinfo
  {author} {\bibfnamefont {S.}~\bibnamefont {Hoedl}}, \bibinfo {author}
  {\bibfnamefont {A.}~\bibnamefont {Lambrecht}}, \bibinfo {author}
  {\bibfnamefont {V.~V.}\ \bibnamefont {Nesvizhevsky}}, \bibinfo {author}
  {\bibfnamefont {G.}~\bibnamefont {Pignol}}, \bibinfo {author} {\bibfnamefont
  {K.~V.}\ \bibnamefont {Protasov}}, \bibinfo {author} {\bibfnamefont
  {S.}~\bibnamefont {Reynaud}},\ and\ \bibinfo {author} {\bibfnamefont
  {Y.}~\bibnamefont {Sobolev}},\ }\bibfield  {title} {\bibinfo {title}
  {Short-range fundamental forces},\ }\href
  {https://doi.org/https://doi.org/10.1016/j.crhy.2011.05.004} {\bibfield
  {journal} {\bibinfo  {journal} {Comptes Rendus Physique}\ }\textbf {\bibinfo
  {volume} {12}},\ \bibinfo {pages} {755} (\bibinfo {year} {2011})}\BibitemShut
  {NoStop}%
\bibitem [{\citenamefont {Bae\ss{}ler}\ \emph {et~al.}(2007)\citenamefont
  {Bae\ss{}ler}, \citenamefont {Nesvizhevsky}, \citenamefont {Protasov},\ and\
  \citenamefont {Voronin}}]{Baeler_2007}%
  \BibitemOpen
  \bibfield  {author} {\bibinfo {author} {\bibfnamefont {S.}~\bibnamefont
  {Bae\ss{}ler}}, \bibinfo {author} {\bibfnamefont {V.~V.}\ \bibnamefont
  {Nesvizhevsky}}, \bibinfo {author} {\bibfnamefont {K.~V.}\ \bibnamefont
  {Protasov}},\ and\ \bibinfo {author} {\bibfnamefont {A.~Y.}\ \bibnamefont
  {Voronin}},\ }\bibfield  {title} {\bibinfo {title} {Constraint on the
  coupling of axionlike particles to matter via an ultracold neutron
  gravitational experiment},\ }\href
  {https://doi.org/10.1103/PhysRevD.75.075006} {\bibfield  {journal} {\bibinfo
  {journal} {Phys. Rev. D}\ }\textbf {\bibinfo {volume} {75}},\ \bibinfo
  {pages} {075006} (\bibinfo {year} {2007})}\BibitemShut {NoStop}%
\bibitem [{\citenamefont {Borisov}\ \emph {et~al.}(1988)\citenamefont
  {Borisov}, \citenamefont {Borovikava}, \citenamefont {Vasil‘ev},
  \citenamefont {Grigor‘eva}, \citenamefont {Ivanov}, \citenamefont
  {Kashukeev}, \citenamefont {Nesvizhevskii}, \citenamefont {Serebrov},\ and\
  \citenamefont {Yaidzhiev}}]{Borisov_1988}%
  \BibitemOpen
  \bibfield  {author} {\bibinfo {author} {\bibfnamefont {Y.~V.}\ \bibnamefont
  {Borisov}}, \bibinfo {author} {\bibfnamefont {N.~V.}\ \bibnamefont
  {Borovikava}}, \bibinfo {author} {\bibfnamefont {A.~V.}\ \bibnamefont
  {Vasil‘ev}}, \bibinfo {author} {\bibfnamefont {L.~A.}\ \bibnamefont
  {Grigor‘eva}}, \bibinfo {author} {\bibfnamefont {S.~N.}\ \bibnamefont
  {Ivanov}}, \bibinfo {author} {\bibfnamefont {N.~T.}\ \bibnamefont
  {Kashukeev}}, \bibinfo {author} {\bibfnamefont {V.~V.}\ \bibnamefont
  {Nesvizhevskii}}, \bibinfo {author} {\bibfnamefont {A.~P.}\ \bibnamefont
  {Serebrov}},\ and\ \bibinfo {author} {\bibfnamefont {P.~S.}\ \bibnamefont
  {Yaidzhiev}},\ }\bibfield  {title} {\bibinfo {title} {On the feasibility of
  using ultracold neutrons to measure the electric charge of the neutron},\
  }\href@noop {} {\bibfield  {journal} {\bibinfo  {journal} {Sov. Phys. Tech.
  Phys.}\ }\textbf {\bibinfo {volume} {33}},\ \bibinfo {pages} {574} (\bibinfo
  {year} {1988})}\BibitemShut {NoStop}%
\bibitem [{\citenamefont {Mart\'{\i}n-Ruiz}\ \emph
  {et~al.}(2015{\natexlab{b}})\citenamefont {Mart\'{\i}n-Ruiz}, \citenamefont
  {Frank},\ and\ \citenamefont {Urrutia}}]{Martin_2015}%
  \BibitemOpen
  \bibfield  {author} {\bibinfo {author} {\bibfnamefont {A.}~\bibnamefont
  {Mart\'{\i}n-Ruiz}}, \bibinfo {author} {\bibfnamefont {A.}~\bibnamefont
  {Frank}},\ and\ \bibinfo {author} {\bibfnamefont {L.~F.}\ \bibnamefont
  {Urrutia}},\ }\bibfield  {title} {\bibinfo {title} {Analysis of the quantum
  bouncer using polymer quantization},\ }\href
  {https://doi.org/10.1103/PhysRevD.92.045018} {\bibfield  {journal} {\bibinfo
  {journal} {Phys. Rev. D}\ }\textbf {\bibinfo {volume} {92}},\ \bibinfo
  {pages} {045018} (\bibinfo {year} {2015}{\natexlab{b}})}\BibitemShut
  {NoStop}%
\bibitem [{\citenamefont {Abele}\ \emph {et~al.}(2010)\citenamefont {Abele},
  \citenamefont {Jenke}, \citenamefont {Leeb},\ and\ \citenamefont
  {Schmiedmayer}}]{Abele_2010}%
  \BibitemOpen
  \bibfield  {author} {\bibinfo {author} {\bibfnamefont {H.}~\bibnamefont
  {Abele}}, \bibinfo {author} {\bibfnamefont {T.}~\bibnamefont {Jenke}},
  \bibinfo {author} {\bibfnamefont {H.}~\bibnamefont {Leeb}},\ and\ \bibinfo
  {author} {\bibfnamefont {J.}~\bibnamefont {Schmiedmayer}},\ }\bibfield
  {title} {\bibinfo {title} {Ramsey's method of separated oscillating fields
  and its application to gravitationally induced quantum phase shifts},\ }\href
  {https://doi.org/10.1103/PhysRevD.81.065019} {\bibfield  {journal} {\bibinfo
  {journal} {Phys. Rev. D}\ }\textbf {\bibinfo {volume} {81}},\ \bibinfo
  {pages} {065019} (\bibinfo {year} {2010})}\BibitemShut {NoStop}%
\bibitem [{\citenamefont {Kosteleck\'y}\ and\ \citenamefont
  {Mewes}(2002{\natexlab{b}})}]{Kostelecky_2002}%
  \BibitemOpen
  \bibfield  {author} {\bibinfo {author} {\bibfnamefont {V.~A.}\ \bibnamefont
  {Kosteleck\'y}}\ and\ \bibinfo {author} {\bibfnamefont {M.}~\bibnamefont
  {Mewes}},\ }\bibfield  {title} {\bibinfo {title} {Signals for {L}orentz
  violation in electrodynamics},\ }\href
  {https://doi.org/10.1103/PhysRevD.66.056005} {\bibfield  {journal} {\bibinfo
  {journal} {Phys. Rev. D}\ }\textbf {\bibinfo {volume} {66}},\ \bibinfo
  {pages} {056005} (\bibinfo {year} {2002}{\natexlab{b}})}\BibitemShut
  {NoStop}%
\bibitem [{\citenamefont {Ding}\ and\ \citenamefont
  {Kosteleck\'y}(2016)}]{Ding_2016}%
  \BibitemOpen
  \bibfield  {author} {\bibinfo {author} {\bibfnamefont {Y.}~\bibnamefont
  {Ding}}\ and\ \bibinfo {author} {\bibfnamefont {V.~A.}\ \bibnamefont
  {Kosteleck\'y}},\ }\bibfield  {title} {\bibinfo {title} {Lorentz-violating
  spinor electrodynamics and {P}enning traps},\ }\href
  {https://doi.org/10.1103/PhysRevD.94.056008} {\bibfield  {journal} {\bibinfo
  {journal} {Phys. Rev. D}\ }\textbf {\bibinfo {volume} {94}},\ \bibinfo
  {pages} {056008} (\bibinfo {year} {2016})}\BibitemShut {NoStop}%
\bibitem [{\citenamefont {Bertolami}(1986)}]{Bertolami_1986}%
  \BibitemOpen
  \bibfield  {author} {\bibinfo {author} {\bibfnamefont {O.}~\bibnamefont
  {Bertolami}},\ }\bibfield  {title} {\bibinfo {title} {Testing the baryon
  number or hypercharge interaction with a neutron interferometric device},\
  }\href {https://doi.org/10.1142/S0217732386000476} {\bibfield  {journal}
  {\bibinfo  {journal} {Modern Physics Letters A}\ }\textbf {\bibinfo {volume}
  {01}},\ \bibinfo {pages} {383} (\bibinfo {year} {1986})}\BibitemShut
  {NoStop}%
\bibitem [{\citenamefont {Saha}(2014)}]{Saha_2014}%
  \BibitemOpen
  \bibfield  {author} {\bibinfo {author} {\bibfnamefont {A.}~\bibnamefont
  {Saha}},\ }\bibfield  {title} {\bibinfo {title}
  {Colella-{O}verhauser-{W}erner test of the weak equivalence principle: A
  low-energy window to look into the noncommutative structure of space-time?},\
  }\href {https://doi.org/10.1103/PhysRevD.89.025010} {\bibfield  {journal}
  {\bibinfo  {journal} {Phys. Rev. D}\ }\textbf {\bibinfo {volume} {89}},\
  \bibinfo {pages} {025010} (\bibinfo {year} {2014})}\BibitemShut {NoStop}%
\bibitem [{\citenamefont {Li}\ \emph {et~al.}(2018)\citenamefont {Li},
  \citenamefont {Yi}, \citenamefont {You-Gen}, \citenamefont {Cheng-Zhou},
  \citenamefont {Hong-Sheng},\ and\ \citenamefont {Lan-Fang}}]{Xiang_2018}%
  \BibitemOpen
  \bibfield  {author} {\bibinfo {author} {\bibfnamefont {X.}~\bibnamefont
  {Li}}, \bibinfo {author} {\bibfnamefont {L.}~\bibnamefont {Yi}}, \bibinfo
  {author} {\bibfnamefont {S.}~\bibnamefont {You-Gen}}, \bibinfo {author}
  {\bibfnamefont {L.}~\bibnamefont {Cheng-Zhou}}, \bibinfo {author}
  {\bibfnamefont {H.}~\bibnamefont {Hong-Sheng}},\ and\ \bibinfo {author}
  {\bibfnamefont {X.}~\bibnamefont {Lan-Fang}},\ }\bibfield  {title} {\bibinfo
  {title} {Generalized uncertainty principles, effective {N}ewton constant and
  the regular black hole},\ }\href
  {https://doi.org/https://doi.org/10.1016/j.aop.2018.07.021} {\bibfield
  {journal} {\bibinfo  {journal} {Annals of Physics}\ }\textbf {\bibinfo
  {volume} {396}},\ \bibinfo {pages} {334} (\bibinfo {year}
  {2018})}\BibitemShut {NoStop}%
\bibitem [{\citenamefont {Farahani}\ \emph {et~al.}(2020)\citenamefont
  {Farahani}, \citenamefont {Hassanabadi}, \citenamefont {Kříž},
  \citenamefont {Chung},\ and\ \citenamefont {Zarrinkamar}}]{Farahani_2020}%
  \BibitemOpen
  \bibfield  {author} {\bibinfo {author} {\bibfnamefont {N.}~\bibnamefont
  {Farahani}}, \bibinfo {author} {\bibfnamefont {H.}~\bibnamefont
  {Hassanabadi}}, \bibinfo {author} {\bibfnamefont {J.}~\bibnamefont
  {Kříž}}, \bibinfo {author} {\bibfnamefont {W.~S.}\ \bibnamefont {Chung}},\
  and\ \bibinfo {author} {\bibfnamefont {S.}~\bibnamefont {Zarrinkamar}},\
  }\bibfield  {title} {\bibinfo {title} {D{SR}-{GUP} black hole based on {COW}
  experiment and {E}instein-{B}ohr’s photon box},\ }\href
  {https://doi.org/10.1140/epjc/s10052-020-8270-6} {\bibfield  {journal}
  {\bibinfo  {journal} {The European Physical Journal C}\ }\textbf {\bibinfo
  {volume} {80}},\ \bibinfo {pages} {696} (\bibinfo {year} {2020})}\BibitemShut
  {NoStop}%
\bibitem [{\citenamefont {Staudenmann}\ \emph {et~al.}(1980)\citenamefont
  {Staudenmann}, \citenamefont {Werner}, \citenamefont {Colella},\ and\
  \citenamefont {Overhauser}}]{COW_1980}%
  \BibitemOpen
  \bibfield  {author} {\bibinfo {author} {\bibfnamefont {J.~L.}\ \bibnamefont
  {Staudenmann}}, \bibinfo {author} {\bibfnamefont {S.~A.}\ \bibnamefont
  {Werner}}, \bibinfo {author} {\bibfnamefont {R.}~\bibnamefont {Colella}},\
  and\ \bibinfo {author} {\bibfnamefont {A.~W.}\ \bibnamefont {Overhauser}},\
  }\bibfield  {title} {\bibinfo {title} {Gravity and inertia in quantum
  mechanics},\ }\href {https://doi.org/10.1103/PhysRevA.21.1419} {\bibfield
  {journal} {\bibinfo  {journal} {Phys. Rev. A}\ }\textbf {\bibinfo {volume}
  {21}},\ \bibinfo {pages} {1419} (\bibinfo {year} {1980})}\BibitemShut
  {NoStop}%
\bibitem [{\citenamefont {Bonse}\ and\ \citenamefont
  {Wroblewski}(1983)}]{Bonse_1983}%
  \BibitemOpen
  \bibfield  {author} {\bibinfo {author} {\bibfnamefont {U.}~\bibnamefont
  {Bonse}}\ and\ \bibinfo {author} {\bibfnamefont {T.}~\bibnamefont
  {Wroblewski}},\ }\bibfield  {title} {\bibinfo {title} {Measurement of neutron
  quantum interference in noninertial frames},\ }\href
  {https://doi.org/10.1103/PhysRevLett.51.1401} {\bibfield  {journal} {\bibinfo
   {journal} {Phys. Rev. Lett.}\ }\textbf {\bibinfo {volume} {51}},\ \bibinfo
  {pages} {1401} (\bibinfo {year} {1983})}\BibitemShut {NoStop}%
\bibitem [{\citenamefont {Bonse}\ and\ \citenamefont
  {Wroblewski}(1984)}]{Bonse_1984}%
  \BibitemOpen
  \bibfield  {author} {\bibinfo {author} {\bibfnamefont {U.}~\bibnamefont
  {Bonse}}\ and\ \bibinfo {author} {\bibfnamefont {T.}~\bibnamefont
  {Wroblewski}},\ }\bibfield  {title} {\bibinfo {title} {Dynamical diffraction
  effects in noninertial neutron interferometry},\ }\href
  {https://doi.org/10.1103/PhysRevD.30.1214} {\bibfield  {journal} {\bibinfo
  {journal} {Phys. Rev. D}\ }\textbf {\bibinfo {volume} {30}},\ \bibinfo
  {pages} {1214} (\bibinfo {year} {1984})}\BibitemShut {NoStop}%
\bibitem [{\citenamefont {Werner}\ \emph {et~al.}(1988)\citenamefont {Werner},
  \citenamefont {Kaiser}, \citenamefont {Arif},\ and\ \citenamefont
  {Clothier}}]{Werner_1988}%
  \BibitemOpen
  \bibfield  {author} {\bibinfo {author} {\bibfnamefont {S.}~\bibnamefont
  {Werner}}, \bibinfo {author} {\bibfnamefont {H.}~\bibnamefont {Kaiser}},
  \bibinfo {author} {\bibfnamefont {M.}~\bibnamefont {Arif}},\ and\ \bibinfo
  {author} {\bibfnamefont {R.}~\bibnamefont {Clothier}},\ }\bibfield  {title}
  {\bibinfo {title} {Neutron interference induced by gravity: New results and
  interpretations},\ }\href
  {https://doi.org/https://doi.org/10.1016/0378-4363(88)90141-6} {\bibfield
  {journal} {\bibinfo  {journal} {Physica B+C}\ }\textbf {\bibinfo {volume}
  {151}},\ \bibinfo {pages} {22} (\bibinfo {year} {1988})}\BibitemShut
  {NoStop}%
\bibitem [{\citenamefont {Littrell}\ \emph {et~al.}(1997)\citenamefont
  {Littrell}, \citenamefont {Allman},\ and\ \citenamefont
  {Werner}}]{Littrell_1997}%
  \BibitemOpen
  \bibfield  {author} {\bibinfo {author} {\bibfnamefont {K.~C.}\ \bibnamefont
  {Littrell}}, \bibinfo {author} {\bibfnamefont {B.~E.}\ \bibnamefont
  {Allman}},\ and\ \bibinfo {author} {\bibfnamefont {S.~A.}\ \bibnamefont
  {Werner}},\ }\bibfield  {title} {\bibinfo {title} {Two-wavelength-difference
  measurement of gravitationally induced quantum interference phases},\ }\href
  {https://doi.org/10.1103/PhysRevA.56.1767} {\bibfield  {journal} {\bibinfo
  {journal} {Phys. Rev. A}\ }\textbf {\bibinfo {volume} {56}},\ \bibinfo
  {pages} {1767} (\bibinfo {year} {1997})}\BibitemShut {NoStop}%
\end{thebibliography}%

\end{document}